\documentclass[prl,reprint]{revtex4-1}
\usepackage{graphicx,url}
\usepackage{amsmath,braket}

\DeclareMathOperator{\csch}{csch}
\renewcommand{\vec}[1]{\mathbf{#1}}
\newcommand{\kB}{k_{\mathrm{B}}}

\begin{document}
\title{Evaporative cooling  to a Rydberg crystal close to its ground state}
\author{M.~Brune${}^1$ and D.J.~Papoular${}^2$}
\affiliation{
  \centerline{{${}^1$Laboratoire Kastler Brossel, Coll\`ege  de  France,  CNRS,
  ENS-Universit\'e  PSL, Sorbonne  Universit\'e, France}}
  ${}^2$LPTM, UMR 8089 CNRS \& Univ.~Cergy--Pontoise, France}
\date{\today}

\begin{abstract}
  We theoretically show how to obtain a long one--dimensional crystal
    near its quantum ground state. We rely on an evaporative cooling scheme
  applicable to many--body systems  
  with nonzero--ranged interactions.
  Despite
  the absence of periodic potentials, the final state is a crystal which
  exhibits long--range spatial order.
    We describe the scheme  thermodynamically, applying the truncated
  Boltzmann distribution  to the collective excitations  of the chain,
  and show that it leads to a novel quasi--equilibrium many--body
  state.
  For longer chains, 
  comprising about $1000$ atoms,
  we emphasize  the  quasi--universality   of  the
  evaporation curve. Such exceptionally long 1D crystals
  are only accessible deep in the quantum regime.
  We perform our analysis on the example  
  of an initially thermal chain of circular Rydberg atoms
  confined to a one--dimensional (1D) geometry. 
  Our scheme may be applied to
  other quantum systems with long--ranged interactions
  such as polar molecules.
\end{abstract}

\maketitle

Systems presenting long--ranged interactions
exhibit strongly--correlated crystalline phases
\cite{wigner:PR1934,grimes:PRL1979,
  bohnet:Science2016,jordan:PRL2019}.
Among them, quantum crystals are those whose constituents undergo large--amplitude
zero--point motion \cite{guyer:SolidStatePhysics1969}.
The collective nature of their excitations leads to spectacular phenomena
including 
the Tkachenko oscillations of a vortex lattice
in a superfluid \cite{sonin:RMP1987,coddington:PRL2003},
the giant plasticity of helium crystals \cite{haziot:PRL2013}, and
supersolidity in ultracold gases presenting
interactions beyond the contact limit
\cite{li:Nature2017,leonard:Nature2017,tanzi:PRL2019,bottcher:PRX2019,chomaz:PRX2019}.

Up to now, the investigation of 1D quantum crystals has been hindered
by the difficulty of obtaining large crystals in this geometry, 
where
thermal and quantum fluctuations both destroy long--range order in macroscopic systems
\cite{mermin:PhysRev1968}. Nevertheless, crystallization does occur in finite--sized systems
\cite{vu:arXiv2019}.
It has been unambiguously observed 
in the absence of any external periodic potential
only in  small systems of up to fifty ions
\cite{jurcevic:Nature2014,richerme:Nature2014,senko:Science2014,lechner:PRA2016}
or ten electrons \cite{deshpande:Nature2008,shapir:Science2019}.
The realization of larger 1D crystals requires going deep
into the quantum regime. There,
thermal fluctuations are suppressed, and long--range order is only
limited by quantum fluctuations, 
which are less stringent \cite{mermin:PhysRev1968}.
The realization of large 1D crystals will pave the way
towards the investigation of 1D quantum crystals, where one may 
look for e.g.\ giant plasticity through the tunneling of defects
\cite{balibar:CRPhys2016,partner:PhysicaB2015}.

We focus on one way of obtaining spatial order which relies on 
strong nonzero--range dipole interactions between Rydberg atoms
\cite{kleppner:SciAm1981}.
Rydberg atoms are ideally suited for quantum information
processing \cite{saffman:RMP2010,zeng:PRL2017}
and quantum simulation \cite{nguyen:PRX2018,weimer:NatPhys2010}.
Nontrivial many--body states
\cite{pohl:PRL2010,schauss:Nature2012,schauss:Science2015}
of up to 50 atoms manipulated with optical tweezers
have been prepared through resonant coupling to Rydberg states
\cite{bernien:Nature2017,barredo:Nature2018,labuhn:Nature2016,leseleuc:Science2019}.
Rydberg states may be weakly admixed to the atomic ground state
\cite{pupillo:PRL2010,jau:NatPhys3016,zeiher:NatPhys2016}
or resonantly excited \cite{low:JPB2012} so as to
study
the interplay between anisotropic interactions and disorder
or frustration
\cite{glaetzle:PRX2014}.
Quantum gases resonantly coupled
to Rydberg states have been predicted to exhibit a quantum phase
transition to a Rydberg crystal \cite{weimer:PRL2008},
leading to a universal scaling behavior 
observed 
in the critical region \cite{low:PRA2009}.

In all those cases,  low--angular--momentum
Rydberg states were considered, leading to a strong limitation on the
lifetime
($100\,\mathrm{\mu s}$ per atom, a few $\mathrm{\mu s}$
for many atoms), limiting the size of the system.
Circular Rydberg atoms
\cite{hulet:PRL1983,anderson:PRA2013,zhelyazkova:PRA2016,signoles:PRL2017},
whose excited electron has maximal
orbital and magnetic quantum numbers,
overcome this limitation and offer a very promising platform for the quantum simulation
of many--body problems \cite{nguyen:PRX2018}.
Using spontaneous emission inhibition \cite{kleppner:PRL1981,hulet:PRL1985},
their already long lifetime (30 ms) is  expected to be extended to more than
$1\,\mathrm{min}$.
This timescale allows for implementing an evaporative cooling scheme
applicable to Rydberg atoms
\cite{nguyen:PRX2018}, whose classical analysis shows great promise for reaching
extremely low temperatures.

\begin{figure}
  \includegraphics[width=.7\linewidth]
  {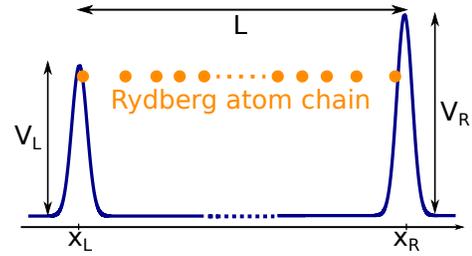}
  \caption{\label{fig:trappedchain}
    A Rydberg atom chain (orange) confined in a 1D trap of size $L$.
    The potential maxima $V_L$ and $V_R$ satisfy $V_L<V_R$, 
    so that atoms are expelled from the left edge of the trap. 
  }
\end{figure}

In this Letter, we
show that 
large 1D Rydberg crystals may be prepared very close to
their quantum ground state 
in realistic experimental conditions \cite{nguyen:PRX2018,cortinas:arXiv2019}
through this evaporative cooling scheme.
Despite the absence of any spatially periodic potential, these
crystals exhibit
long--range spatial order.
This is in  stark contrast to the classical analysis of 1D systems,  which
would predict the absence of long--range order \cite{mermin:PhysRev1968}.
We introduce a quantum thermodynamic model,
applying the truncated Boltzmann distribution to the collective
excitations of the chain.
We show that it leads to a novel
quasi--equilibrium regime 
which differs
from the truncated Bose--Einstein distribution 
applicable to quantum--degenerate
gases \cite{yamashita:PRA1999}.
In contrast to dilute systems where the evaporation is driven by
two--body collisions \cite{luiten:PRA1996}, the mechanism we describe here
hinges on many--body physics, whereby the phonons present in the chain
lead to the expulsion of a single atom. Hence, it is related to the
quantum evaporation of liquid helium
\cite{johnston:PRL1966,anderson:PhysLett1968,dalfovo:PRL1995},
also predicted to affect cold bosonic atoms \cite{papoular:PRA2016}.

We first consider a fixed number $N$ of Rydberg atoms
confined in a 1D trap of fixed size $L$ (see Fig.~\ref{fig:trappedchain}).
We illustrate our model using the parameters
of Ref.~\cite{nguyen:PRX2018}.
The atoms are confined radially
using the  ponderomotive potential \cite{dutta:PRL2000}
induced by a Laguerre--Gaussian laser beam \cite[chap.~2]{allen:IOP2003}.
They are trapped axially between two optical plugs
yielding the potential
$V_T(x)=V_L\exp[-2(x-x_L)^2/w^2]+V_R\exp[-2(x-x_R)^2/w^2]$.
The barrier width and heights are, respectively, $w=30\,\mathrm{\mu m}$,
$V_L/h=3\,\mathrm{MHz}$, and
$V_R/h=4\,\mathrm{MHz}$.
The trap size $L=x_R-x_L$ is slowly decreased from its initial value
so as to induce
successive atomic expulsions, providing the evaporative cooling.
Unlike for gases, the barrier heights remain constant during the whole process.
The atoms
interact via the strongly repulsive van der Waals interaction
$V(x_i,x_j)=C_6/|x_i-x_j|^6$ with $C_6/h=3\,\mathrm{GHz\,\mu m^6}$,
corresponding to ${}^{87}\mathrm{Rb}$ atoms with the
principal quantum number $n=50$.
The equilibrium 
positions $x^0_1,\ldots x^0_N$ are evenly spaced in the bulk of the chain,
but not  on the edges, due to the finite spatial extent of the barriers.
Two neighboring atoms are distant by $l\approx 5\,\mathrm{\mu m}$,
leading to interaction energies
$C_6/l^6
\approx h\cdot 200\,\mathrm{kHz}$.

We describe the atomic vibrations
in terms of a quadratic Hamiltonian:
\begin{equation}
  \label{eq:hamquadratic}
  H=\sum_{k=1}^N \left[
    \frac{\tilde{p}_k^2}{2m}+\frac{1}{2}m\omega_k^2 \tilde{u}_k^2
  \right]
  \text{ with } \tilde{u}_k=\sum_{n=1}^N R_{nk} u_n
  \ .
\end{equation}
In Eq.~\eqref{eq:hamquadratic},
the $N$ vibrational modes $\{\tilde{u}_k\}$
have the frequencies
$\omega_1<\ldots<\omega_N$, and the $\{\tilde{p}_k\}$
are their conjugate momenta.
They are related to the atomic displacements $\{u_n=x_n-x_n^0\}$
through
the orthogonal matrix $R$.
The applicability of Eq.~\eqref{eq:hamquadratic} only requires
local order \cite[Sec.~I]{SuppMat:2019}:
the averages
$\braket{(u_{n+1}-u_n)^2}^{1/2}$ involving two neighboring atoms
should remain small
compared to 
$l=L/N$. For a thermal chain
at the temperature $T$, this requires
$\kB T< 2C_6/l^6$,
and is well 
satisfied
for up to $1000$ atoms with $l\sim 5\,\mathrm{\mu m}$
and
$\kB T\lesssim h\, 100\,\mathrm{kHz}\approx \kB\, 5\,\mathrm{\mu K}$.

\begin{figure}
  \includegraphics[angle=-90,width=.9\linewidth]
  {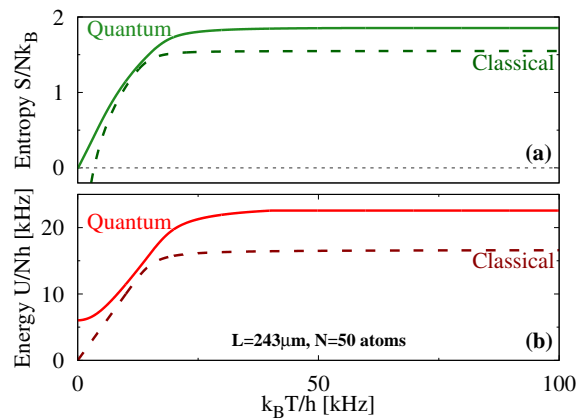}
  \caption{\label{fig:thermoGaussianPlugs}
    \textit{(a)} Entropy 
    and \textit{(b}) energy 
    per particle 
    for $N=50$ atoms in a trap of size $L=243\,\mathrm{\mu m}$
    (close to the end of the evaporation for the chain of
    Fig.~\ref{fig:evapcurveQuantum}).
    The solid (dashed) lines show the quantum (classical) prediction.
  }
\end{figure}
\textit{Classical thermodynamics.}
For a given configuration characterized by the phonon mode energies
$\{\epsilon_k\}_{1\leq k\leq N}$ and 
phases
$\{\phi_k\}_{1\leq k\leq N}$,
the position of the leftmost atom
at time $t$ is
$u_1(t)=\sum_k R_{1k}(2\epsilon_k/(m\omega_k^2))^{1/2}\cos(\omega_k t+\phi_k)$.
It remains trapped 
as long as $|u_1(t)|<u_M$, where $u_M=x_1^0-x_L$.
We consider the time--averaged mean--square displacement 
$\braket{u_1^2}=u_M^2\sum_{k=1}^N\epsilon_k/E_{Mk}$, where
the quantities $E_{Mk}=m\omega_k^2u_M^2/R^2_{1k}$
increase with $k$.
Hence, for a given $\alpha$,
the lowest--energy configurations 
for which $\braket{u_1^2}^{1/2}=\alpha u_M$
are those where only the mode $k=1$ is excited,
with the energy $E=\epsilon_1=\alpha^2 E_{M1}$.
For $\alpha=1/\sqrt{2}$, they correspond
to atom 1 barely reaching $u_1=-u_M$,
i.e.\ to the lowest--energy untrapped configurations.
Their energy $E_M^\mathrm{cl}=m\omega_1^2u_M^2/(2R_{11}^2)$ is set by
$\omega_1$.
Furthermore, numerical simulations of the classical (cl) dynamics of the atom chain
\cite{brune:ClassDynNumerics2017}
have shown the atomic motion
to be chaotic.
Hence, exploiting ergodicity,
the trapped configurations are those with $E<E_M^\mathrm{cl}$.
We describe the quasi--equilibrium thermodynamics of the chain 
using a Boltzmann distribution truncated at the energy $E_M^\mathrm{cl}$,
whose partition function reads:
\begin{equation}
  \label{eq:partfuncl}
  Z^\mathrm{cl}=
  \int_{E<E_M^\mathrm{cl}}
  \frac{\prod [d\tilde{p}_kd\tilde{u}_k]}{h^N}\, e^{-\beta E}
  =\frac{P(N,\beta E_M^\mathrm{cl})}{\beta^N\hbar\omega_1\ldots\hbar\omega_N}
  \ .
\end{equation}
In Eq.~\eqref{eq:partfuncl}, $\beta=1/(\kB T)$ is the inverse temperature,
$E=H(\{\tilde{p}_k,\tilde{u}_k\})$ and
$P(a,z)=\gamma(a,z)/\Gamma(N)$ is the
normalized lower incomplete gamma function \cite{NIST:DLMF}.
The mean (quadratic) energy $U^\mathrm{cl}(L,T)$
associated with the Hamiltonian
$H$ and the entropy $S^\mathrm{cl}(L,T)$
follow from 
$U^\mathrm{cl}=-\partial_\beta\log Z^\mathrm{cl}$ and
$S^\mathrm{cl}/\kB=\log Z^\mathrm{cl} -\beta\partial_\beta\log Z^\mathrm{cl}$.

The function $P(a,z)$ also appears in the thermodynamics
of the evaporation of a gas 
($a=3$ for a harmonic trap) \cite{luiten:PRA1996}.
Here, $a=N$ ranges from $40$ to $1000$,
so that the role of truncation
is strongly enhanced with respect to  gases of ground--state atoms
\cite[Sec.~III]{SuppMat:2019}.
It is important
for $\kB T\gtrsim E_M^\mathrm{cl}/N$.
For larger $T$, all trapped configurations are equally populated.
The probability density for a configuration to have the energy $E$ 
is $NE^{N-1}/(E_M^\mathrm{cl})^N$, hence, nearly all configurations have
energies $\sim E_M^\mathrm{cl}$.
Both $U^\mathrm{cl}$ and $S^\mathrm{cl}$
reach finite maxima 
$U^N_{\mathrm{max}}(L)$ and $S^N_{\mathrm{max}}(L)$
(see Fig.~\ref{fig:thermoGaussianPlugs}), where:
\begin{equation}
  \label{eq:UmaxSmax}
  U^{(N)}_{\mathrm{max}}=\frac{NE_M^\mathrm{cl}}{N+1}
   \text{ and }
  S^{(N)}_{\mathrm{max}}=\kB
  \log\left(
    \frac{(E_M^\mathrm{cl})^N/N!}{\hbar\omega_1\ldots\hbar\omega_N}
  \right)
  .
\end{equation}
For fixed $N$, both maxima
increase with $L$, because less stringent traps
will accommodate higher--energy excitations.
This novel
regime is inaccessible with gases, where an atom whose energy is close
to the evaporation threshold is expelled 
when it undergoes
a collision \cite{walraven:bristol1996,luiten:PRA1996}.
However, it is accessible for a Rydberg chain
\cite[Sec.~II]{SuppMat:2019}.

\begin{figure}
  \includegraphics[angle=-90,width=.9\linewidth]
  {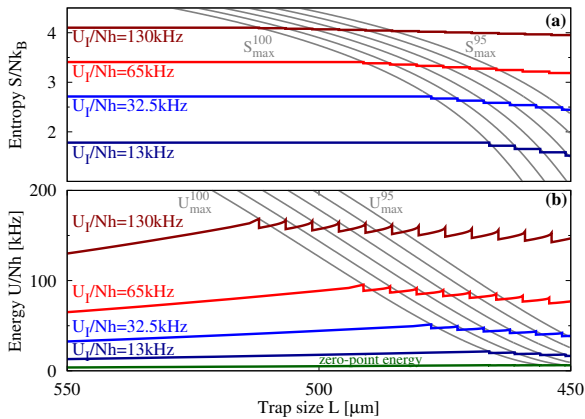}
  \caption{ \label{fig:compare_kin_zoom}
    The first few expulsions for a chain with $N_I=100$ atoms and
    $L_I=550\,\mathrm{\mu m}$,
    in terms of \textit{(a)} entropy $S/N$ and \textit{(b)}
    energy $U/N$ per particle,
    for various initial energies.
    The maxima $S^{(N)}_\mathrm{max}(L)/N$ and $U^{(N)}_\mathrm{max}(L)/N$ are
    shown in gray.
    Each expulsion yields a discontinuity in both $S$ and $U$.
  }
\end{figure}

\textit{Quantum thermodynamics.}
For lower quadratic energies,
we 
use a quantum (quant) description.
Assuming ergodicity in the quantum regime, we
introduce the energy $E_{\vec{n}}=\sum_{k=1}^N\hbar\omega_k(n_k+1/2)$
of the configuration labeled by the integer multiplet
$\vec{n}=\{n_k\}_{1\leq k\leq N}$.
The threshold energy
$E_M^\mathrm{quant}$ for trapped configurations
satisfies:
\begin{equation}
  \label{eq:eMqu}
  E_M^\mathrm{quant}=
  \min_{\vec{n}}
  \left[ E_{\vec{n}}
    \text{ with }
    \sum_{k=1}^N\frac{\hbar\omega_k(n_k+1/2)}{E_{Mk}}\geq \alpha^2 
  \right]
  ,
\end{equation}
where we choose
$\alpha=1/\sqrt{2}$ as in the classical case.
The energy $E_M^\mathrm{quant}$
exceeds both $E_M^\mathrm{cl}$ and
the zero--point energy $E_\mathrm{ZP}=\sum_{k=1}^N\hbar\omega_k/2$.
The quantum partition function 
reads
$Z^\mathrm{quant}=
\sum_{\vec{n}} e^{-\beta E_{\vec{n}}}\: \Theta(E_M^\mathrm{quant}-E_{\vec{n}})$,
where $\Theta$ is the Heaviside function, 
illustrating an important difference
with respect to gases of ground--state atoms.
There,
the truncation selects the trapped single--particle modes
without constraining their populations, 
yielding a truncated Bose--Einstein distribution \cite{yamashita:PRA1999}.
Instead, for Rydberg chains,
the truncation involves 
the 
configuration energies $E_\vec{n}$.
This prevents $Z^\mathrm{quant}$ from factorizing and reflects the
correlations between the trapped phonon modes,
leading to a
novel 
quasi--equilibrium state
which does not obey
a truncated Bose--Einstein distribution.

We assume $E_M^\mathrm{quant}\gg E_\mathrm{ZP}+ \hbar\omega_N$,
which is well satisfied for all parameters considered in this paper.
Then,
$E_M^\mathrm{quant}\approx E_M^\mathrm{cl}+E_\mathrm{ZP}$.
For $\kB T\gtrsim E_M^\mathrm{quant}/N$,
we evaluate the quantum energy $U^\mathrm{quant}(L,T)$ and
entropy $S^\mathrm{quant}(L,T)$ \cite[Sec.~III]{SuppMat:2019}
starting from Eq.~\eqref{eq:partfuncl},
replacing $E_M^\mathrm{cl}$ by $E_M^\mathrm{quant}$
and including
the leading quantum
correction,
proportional to $\hbar^2$ \cite[\S 33]{landau5:Elsevier1980}.
For $\kB T< E_M^\mathrm{quant}/N$,
the energy and entropy 
reflect the non--truncated thermodynamics
of a harmonic oscillator chain.
They overlap with $U^\mathrm{quant},S^\mathrm{quant}$
for a range of values of $T$,
yielding the
full quantum thermodynamic functions
(see Fig.~\ref{fig:thermoGaussianPlugs}).

\begin{figure}
  \includegraphics[angle=-90,width=.9\linewidth]
  {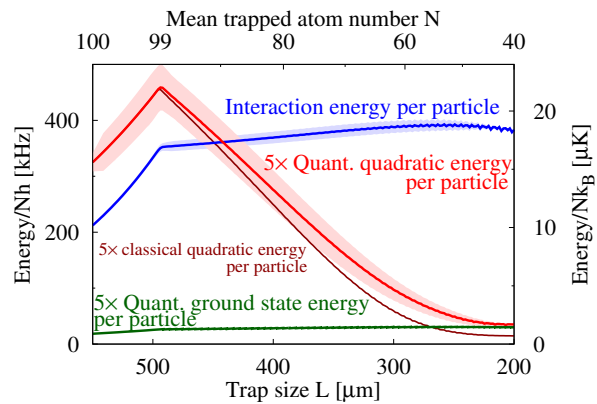}
  \caption{\label{fig:evapcurveQuantum}
    Classical (dark red) and quantum (red)
    predictions for the mean quadratic energy,
    interaction (blue) and ground state (green) energies
    per particle during the evaporation, starting from
    $N_I=100$, $L_I=550\,\mathrm{\mu m}$, down to
    $N_F=40$, $L_F=200\,\mathrm{\mu m}$.
    The shaded red and blue areas show the standard deviations on
    the quadratic and interaction energies.
    Energies are measured in kHz, with
    $h\cdot 100\,\mathrm{kHz}\sim\kB\cdot 5\,\mathrm{\mu K}$.}
\end{figure}

\emph{Evaporation.}
We now describe the evaporation process. 
Initially, the chain comprises
$N=N_I$ atoms in a trap of size $L^{(N)}=L_I$,
with the energy $U^{(N)}=U_I$.
For all considered parameters,
$U_I\gg E_\mathrm{ZP}$, signalling the
classical regime, and $U_I\ll E_{MI}^\mathrm{cl}$,
so that it is described by non--truncated thermodynamics.
Thus, $U_I/N_I=\kB T_I$ is the initial temperature.  
We adiabatically compress the chain by slowly decreasing $L$
(see Fig.~\ref{fig:compare_kin_zoom}).
Hence, the entropy $S^{(N)}$  remains constant.
Expelling an atom is irreversible,
therefore $N$ also remains constant.
However,
$T$ and $U^{(N)}$ increase, whereas
$U^{(N)}_\mathrm{max}(L)$ and $S^\mathrm{(N)}_\mathrm{max}(L)$
decrease. The compression proceeds
until the trap no longer accommodates the entropy,
i.e.\ up to the trap size $L_f^{(N)}$ such that
$S^{(N)}_{\mathrm{max}}(L_f^{(N)})=S^{(N)}$.
This implies $T\rightarrow\infty$, hence,
$U^{(N)}_f=U^{(N)}_{\mathrm{max}}(L_f^{(N)})$.
At this point, the leftmost atom is expelled from the trap,
its kinetic energy being the barrier height $V_L$.
The $(N-1)$ remaining atoms thermalize
to the new initial energy $U^{(N-1)}_i$, where:
\begin{equation}
  \label{eq:Uinew}
  U^{(N-1)}_i=U^{(N)}_f+V_0^{(N)}-V_0^{(N-1)}-V_L
  \ .
\end{equation}
Here, $V_0^{(N)}$ and $V_0^{(N-1)}$ are the static equilibrium energies for
$N$ and $(N-1)$ atoms 
in a trap of size $L_f^{(N)}$.
Then, adiabatic compression resumes until the next expulsion.

The complete evaporation curve consists of a repeated sequence
of these two steps.
Figure \ref{fig:evapcurveQuantum}
compares our classical (dark red)  and quantum (red) predictions, 
down to the trap size
$L_F=200\,\mathrm{\mu m}$ where 
$E_M^\mathrm{quant}\gtrsim E_\mathrm{ZP}+ \hbar\omega_N$.
The result of our classical model closely matches
the classical--dynamics simulations reported
in Ref.~\cite{nguyen:PRX2018} (Fig.~14, phase II).
Our quantum approach predicts that, starting from $N_I=100$ atoms,
the final state with $N_F=40$ atoms obeys
a Bose--Einstein distribution with 
$U_F/(N_Fh)=7.0\,\mathrm{kHz}$,
slightly above the zero--point energy
$E_\mathrm{ZP}/(N_F h)=5.9\,\mathrm{kHz}$.
The shown average energies 
account for
the uncertainty $\Delta U_I=U_I/\sqrt{N_I}=h\cdot 6.5\,\mathrm{kHz}$
on $U_I$, 
which washes out
their jaggedness due to the expulsions
(Fig.~\ref{fig:compare_kin_zoom})
\cite[Sec.~IV]{SuppMat:2019}.
The final state is in the 1D regime if
the radial confinement frequency
$\omega_\perp/(2\pi)\gg U_F/(N_F h)$.
Smaller values of $\omega_\perp$
will lead to quasi--1D chains exhibiting the `zigzag'
transition observed with ion chains
\cite{birkl:Nature1992,partner:PhysicaB2015}
and in electronic systems \cite{ikegami:PRL2012}.

\begin{figure}
  \includegraphics[angle=-90,width=.9\linewidth]
  {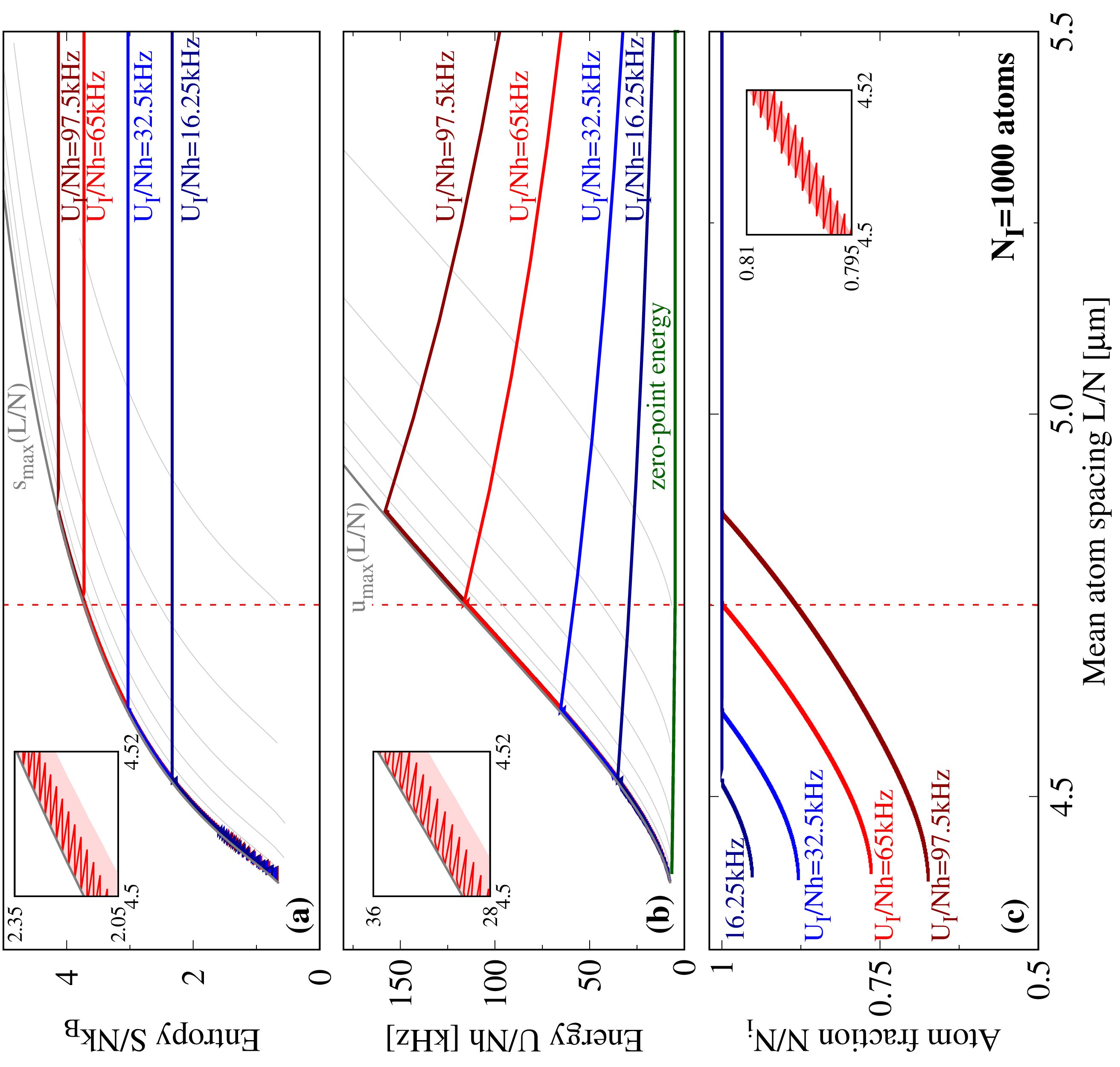}
  \caption{\label{fig:univKSN}
    Quasi--universal evaporation of a chain
    with $N_I=1000$ and 
    $l_I=5.5\,\mathrm{\mu m}$, in terms of the mean
    \textit{(a)} entropy
    and \textit{(b)} energy per particle,
    for various initial energies.
    The thin gray lines
    show $S_\mathrm{max}(L/N)/N$ and $S_\mathrm{max}(L/N)/N$
    for $N=50$, $100$, $200$, $400$, and $800$ (from right to left)
    which converge
    towards $s_\mathrm{max}(l)$ and $u_\mathrm{max}(l)$ (thick gray lines).
    Panel \textit{(c)}: non--universal
    atom fraction $N/N_I$.
    The vertical red line shows the first
    expulsion for $U_I/(N_Ih)=65\,\mathrm{kHz}$.
    The insets zoom in on the same
    small fraction of the curves
    for $U_I/(N_Ih)=65\,\mathrm{kHz}$, and show the jagged 
    curves obtained before averaging;
    the shaded areas show the
    standard deviations.
  }
\end{figure}

\textit{Quasi--universality for longer chains.}
Finally, we focus on long chains with $N_I\approx 1000$,
keeping $l_I=L_I/N_I=5.5\,\mathrm{\mu m}$.
Then, the inhomogeneities near the trap edges 
are negligible, and both
$S/N=s(l,T)$ and $U/N=u(l,T)$ only depend on 
$l=L/N$
and $T$.
The evaporation
is conveniently described in terms of $l$, $s$, $u$, and
the atom number fraction $n=N/N_I$.
The evaporation curve consists of two parts (see Fig.~\ref{fig:univKSN}).
First,
the initial compression at constant $N_I$ depends on
$u_I=U_I/N_I$.
The second part consists of all subsequent expulsions and compressions.
The mean distance $l$ increases at each expulsion and decreases during
each compression; on average, $l$ decreases.
The quantities $s$ and $u$ always remain close to the universal curves
$s_\mathrm{max}(l)=S_\mathrm{max}(N,L)/N$
and $u_\mathrm{max}(l)=U_\mathrm{max}(N,L)/N$,
respectively \cite[Sec.~V]{SuppMat:2019}.
Their fluctuations, 
visible in the insets of Fig.~\ref{fig:univKSN}
for $N_I=1000$ and $\kB T_I/h=65\,\mathrm{kHz}$,
decrease with increasing $N_I$ for two reasons.
First, the changes $\delta u$ and $\delta s$
in the energy and entropy per particle
upon expelling an atom
are decreasing functions of $N$.
Second, larger 
$N_I$ lead to smaller 
$\Delta U_I=U_I/\sqrt{N_I}$
and, hence, to smaller uncertainties on $s$ and $u$.
Quasi--universality also applies to
the fluctuations $\Delta u$
and $\Delta s$ on the energy and entropy
\cite[Sec.~V]{SuppMat:2019}.

The fraction $n=N/N_I$
(Fig.~\ref{fig:univKSN}\textit{(c)})
is not universal
\cite[Fig.~S6]{SuppMat:2019}.
For $N_I=1000$,
$n$ reaches a stationary value $n_F$ as $u$ 
goes to
$e_\mathrm{ZP}(l)=E_\mathrm{ZP}(l)/N$.
The value $n_F(u_I)$
is a decreasing function of $u_I=U_I/N_I$. The curves on Fig.~\ref{fig:univKSN} are truncated at the minimum value
$l=4.4\,\mathrm{\mu m}$ where
$E_M^\mathrm{quant}\gtrsim E_\mathrm{ZP}+\hbar\omega_N$. 
Then, for $\kB T_I/h=65\,\mathrm{kHz}$,
the chain comprises $N_F=764$ atoms
with the energy
$U_F/(N_Fh)=8.5\,\mathrm{kHz}$,
close to 
$E_\mathrm{ZP}/(N_Fh)=6.6\,\mathrm{kHz}$.

The final state of such a  long  chain
is a crystal exhibiting
true long--range order, with all spatial correlators
$C_{nm}=\braket{(u_n-u_m)^2}\ll l^2$ \cite[Fig.~S1]{SuppMat:2019}.
This is only possible deep in the quantum regime,
where
thermal fluctuations are suppressed \cite{mermin:PhysRev1968}.
The crystalline order may be fully characterized experimentally
through microwave spectroscopy,
revealing the regularity and fluctuations of the lattice parameter,
combined with spatially---resolved ground state imaging
\cite{barredo:Science2016,barredo:Nature2018}.

We have introduced a quantum thermodynamic model for
the evaporative cooling of 1D Rydberg atom
chains \cite{nguyen:PRX2018}.
Unlike the evaporative cooling of ground--state atoms, the final temperatures
accessible with our scheme are not of the order of the barrier heights. Instead,
they are
determined by
the maximum energy $u_\mathrm{max}(l)$ compatible with the trap. This reflects the
many--body character of the evaporation scheme and
leads to final temperatures that are
radically lower than the barrier heights by three orders of magnitude.
We have shown that, under realistic
experimental conditions, this scheme
yields large near--ground--state Rydberg crystal.
The long--range spatial order of these 
1D structures is a feature of  the deep quantum regime.
Our scheme will also apply to other interacting 1D systems such as
polar molecules \cite{demarco:Science2019,anderegg:Science2019}. There,
the nonzero--ranged interaction between the particles is provided
by the dipole--dipole interaction, 
which scales
with $1/r^3$ and may be made purely repulsive in low--dimensional
geometries \cite{baranov:ChemRev2012}. 

\textit{Outlook.} The following directions warrant further investigation.
\textit{(i)} For higher initial temperatures or mean atom spacings, the
initial state is a liquid and Eq.~\eqref{eq:hamquadratic}
does not hold, but our scheme 
will still drive the system towards its crystalline ground state.
\textit{(ii)} For longer chains, a prolonged evaporation
going beyond the regime of Fig.~\ref{fig:univKSN} leads to
$E_M^\mathrm{quant}\lesssim E_\mathrm{ZP}+\hbar\omega_N$,
in which case the calculation
of the quantum thermodynamic functions is more involved.
\textit{(iii)} The timescale ensuring adiabaticity is set by the
anharmonic processes neglected in Eq.~\eqref{eq:hamquadratic}.
\textit{(iv)} Our scheme is also applicable in 2D,
where the expected ground state 
is a hexagonal crystal
which we shall investigate both
theoretically and experimentally.

\begin{acknowledgments}
  We acknowledge stimulating discussions with
  J.M.~Raimond, Ph.~Lecheminant,
  Y.~Castin, T.~Huillet, L.P.~Pitaevskii,
  C.~Sayrin, and G.V.~Shlyapnikov.
\end{acknowledgments}

%\bibliography{rydbergevapbib}


%%%%%%%%%%%%%%%%%%%%%%%%%%%%%%%%%%%%%%%%%%%%%%%%%%%%%%%%%%%%%%%%%%%%%%%%%%%%%%%%

\clearpage

\centerline{\Large \textbf{Supplemental material}}

\renewcommand{\vec}[1]{\mathbf{#1}}

\setcounter{equation}{0}
\setcounter{figure}{0}
\setcounter{table}{0}
\setcounter{page}{1}
\renewcommand{\theequation}{S\arabic{equation}}
\renewcommand{\thefigure}{S\arabic{figure}}
\renewcommand{\bibnumfmt}[1]{[S#1]}
\renewcommand{\citenumfont}[1]{S#1}

%\begin{document}
%\title{Evaporative cooling  of a Rydberg atom chain to near its ground %state\\
%  \underline{Supplemental Material}}
%\author{M.~Brune${}^1$ and D.J.~Papoular${}^2$}
%\affiliation{
%  \centerline{${}^1$Laboratoire Kastler Brossel,
%    Coll\`ege  de  France,  CNRS,
%    ENS-Universit\'e  PSL, Sorbonne  Universit\'e, France}
%  {${}^2$LPTM,
%    UMR 8089 CNRS \& Univ.~Cergy--Pontoise, France}
%}
%\date{\today}

%\begin{abstract}
  This document provides complementary information on the following topics:
  I.~the applicability of the quadratic Hamiltonian;
  II.~the anharmonic terms and their twofold role;
  III.~the partition function and its numerical evaluation;
  IV.~the observability
  of the adiabatic plateaux with constant atom numbers;
  V.~the quasi--universal description for long chains and its limits.
%\end{abstract}

%\maketitle

\begin{figure*}
  \begin{minipage}{.3\textwidth}
    \includegraphics[angle=-90,width=\textwidth]
    {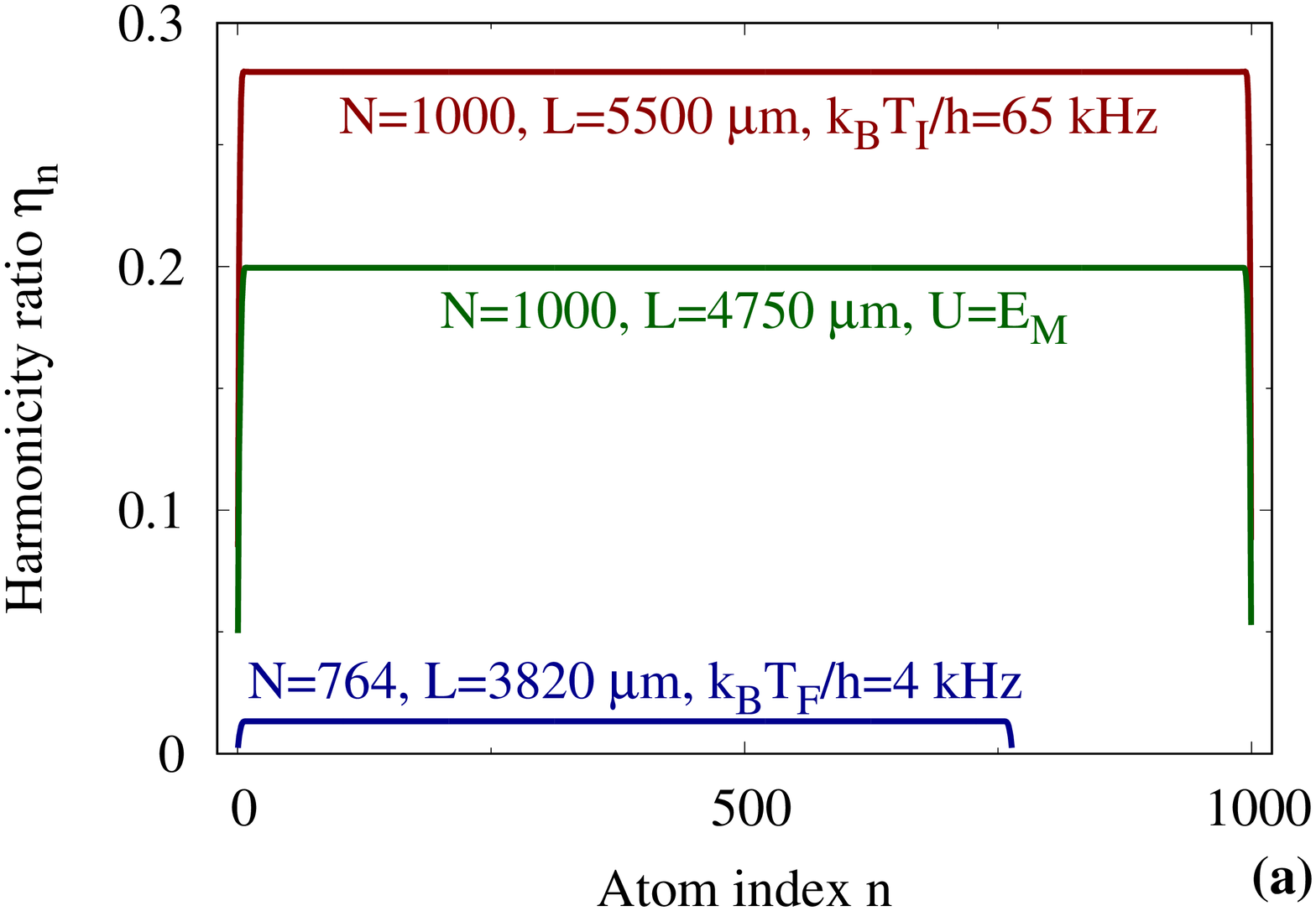}
  \end{minipage}
  \begin{minipage}{.33\textwidth}
    \includegraphics[angle=-90,width=\textwidth]
    {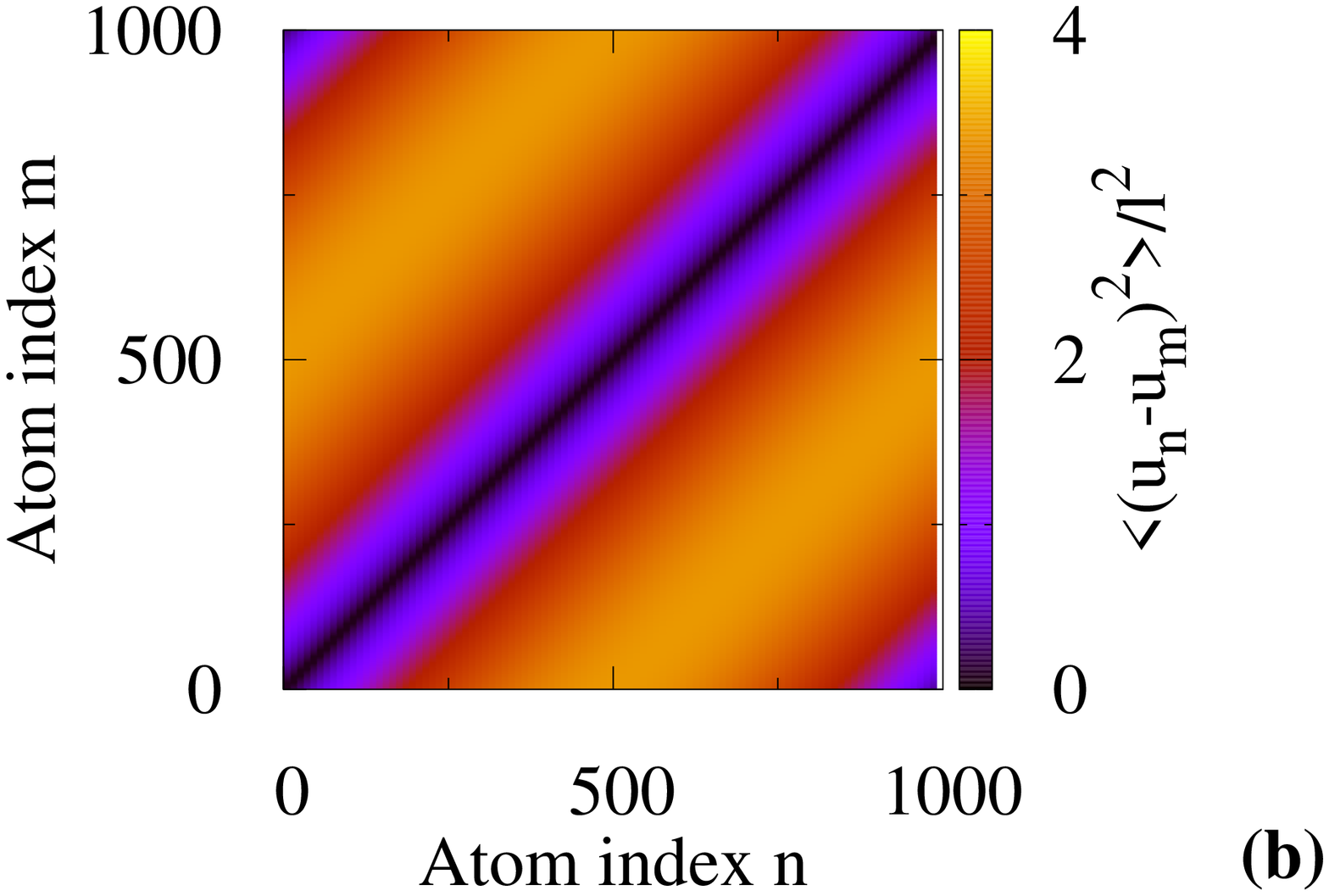}
  \end{minipage}
  \begin{minipage}{.33\textwidth}
    \includegraphics[angle=-90,width=\textwidth]
    {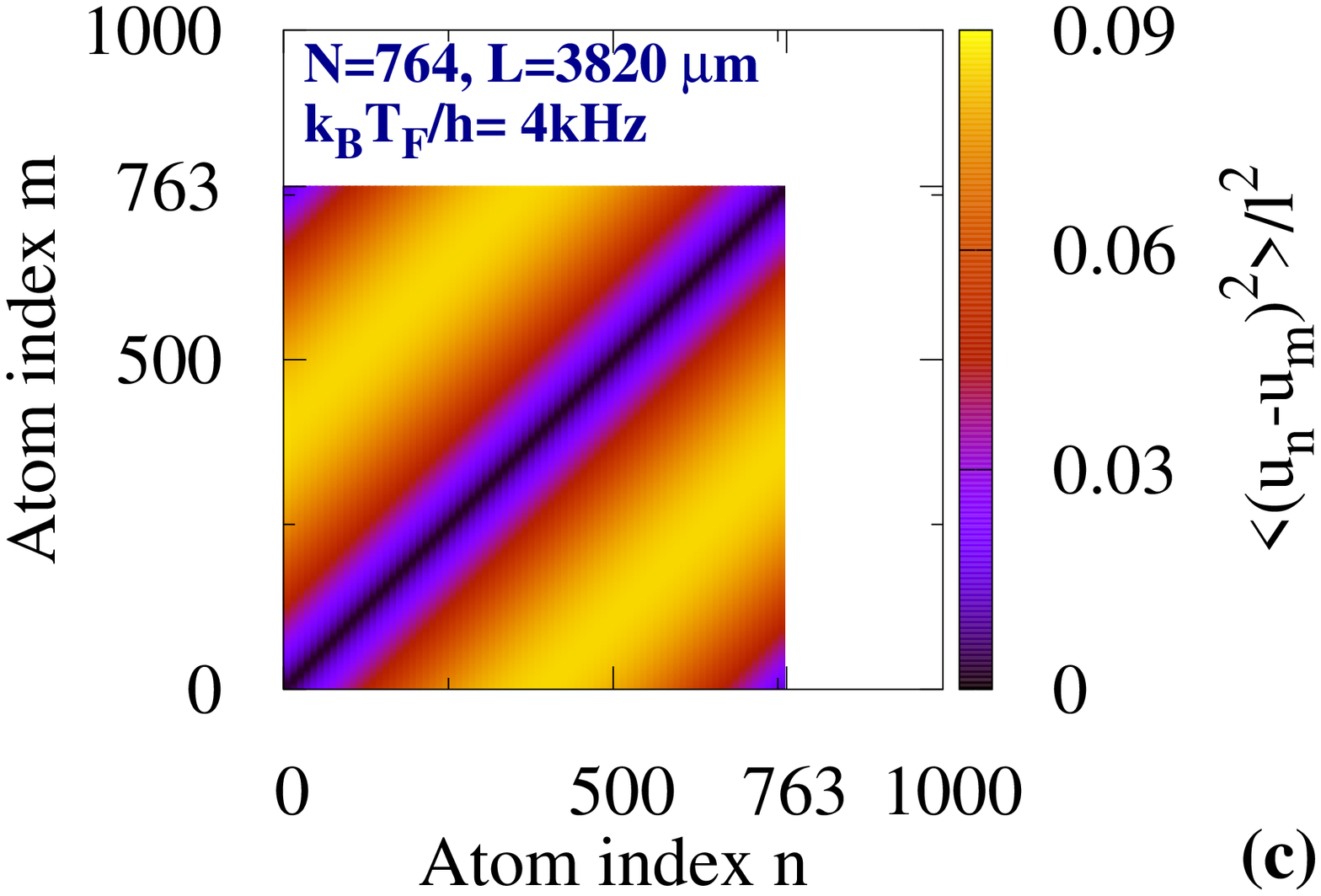}
  \end{minipage}
  \caption{\label{fig:SM:harmratio}
    \textit{(a)} Harmonicity ratio $\eta_n$ for the long chain
    (Fig.~5 in the main text),
    at the beginning of the
    evaporation (red),
    just before the first expulsion (green),
    and at the end of the evaporation (blue).
    \textit{(b)} and \textit{(c)}: Spatial correlator
    $\braket{(u_n-u_m)^2}/l^2$ at the beginning
    (\textit{(b)}, $N=1000$, $L=5500\,\mathrm{\mu m}$)
    and the end
    (\textit{(c)}, $N=764$, $L=3820\,\mathrm{\mu m}$) of the evaporation,
    in units of $l=L/N$.
  }
\end{figure*}
\section{The quadratic hamiltonian}

The Hamiltonian describing the harmonic vibrations of the atoms
about their equilibrium positions $\{x_n^0\}$
(Eq.~1 in the main text) is applicable as soon as the chain exhibits
local order.
Indeed, in the chain bulk, the trapping potential is negligible and, within the
nearest--neighbor approximation, the interaction energy of atom $n$ is
$E_n^I=C_6[1/(x_n-x_{n-1})^6+1/(x_n-x_{n+1})^6]$. Here, $x_n=x_n^0+u_n$ is the position
of atom $n$. Expanding $E_n^I$ to second order
in the displacements $\{u_n\}$,
and exploiting the near--translational invariance,
we find that the harmonic approximation is valid if
$\eta_n=21\braket{(u_{n+1}-u_n)^2}/l^2<1$, where $l=L/N$ is the mean interatomic distance and
the average $\braket{(u_{n+1}-u_n)^2}$ is the spatial correlator between two neighboring
atoms. For a thermal distribution, this condition
reduces to
$\kB T< 2C_6/l^6$. 
Accounting for
the trap and the truncated thermodynamics,
we find this criterion
to be well satisfied all along the evaporation for the long chain of Fig.~5 in the main text
(see Fig.~\ref{fig:SM:harmratio}\textit{(a)}).

The present criterion is less stringent than asking
for the chain to be in a crystalline phase. This is especially true in 1D where thermal
fluctuations quickly rule out long--range order \cite{mermin:PhysRev1968}. For example,
the long chain of Fig.~5 exhibits no long--range spatial correlations in its initial state
($N_i=1000$, $l_i=5.5\,\mathrm{\mu m}$,
$\kB T_i/h=65\,\mathrm{kHz}$). This can be seen on Fig.~\ref{fig:SM:harmratio}\textit{(b)}:
the correlator $\braket{(u_n-u_m)^2}/l^2>1$ for distant atoms.
However, our scheme brings the chain close to its quantum ground state, which does exhibit 
long--range correlations ($\braket{(u_n-u_m)^2}/l^2\ll 1$ for all $n$ and $m$,
see Fig.~\ref{fig:SM:harmratio}\textit{(c)}).

\section{Anharmonic effects}
\begin{figure}
    \centering
    \includegraphics[angle=-90,width=.7\linewidth]
    {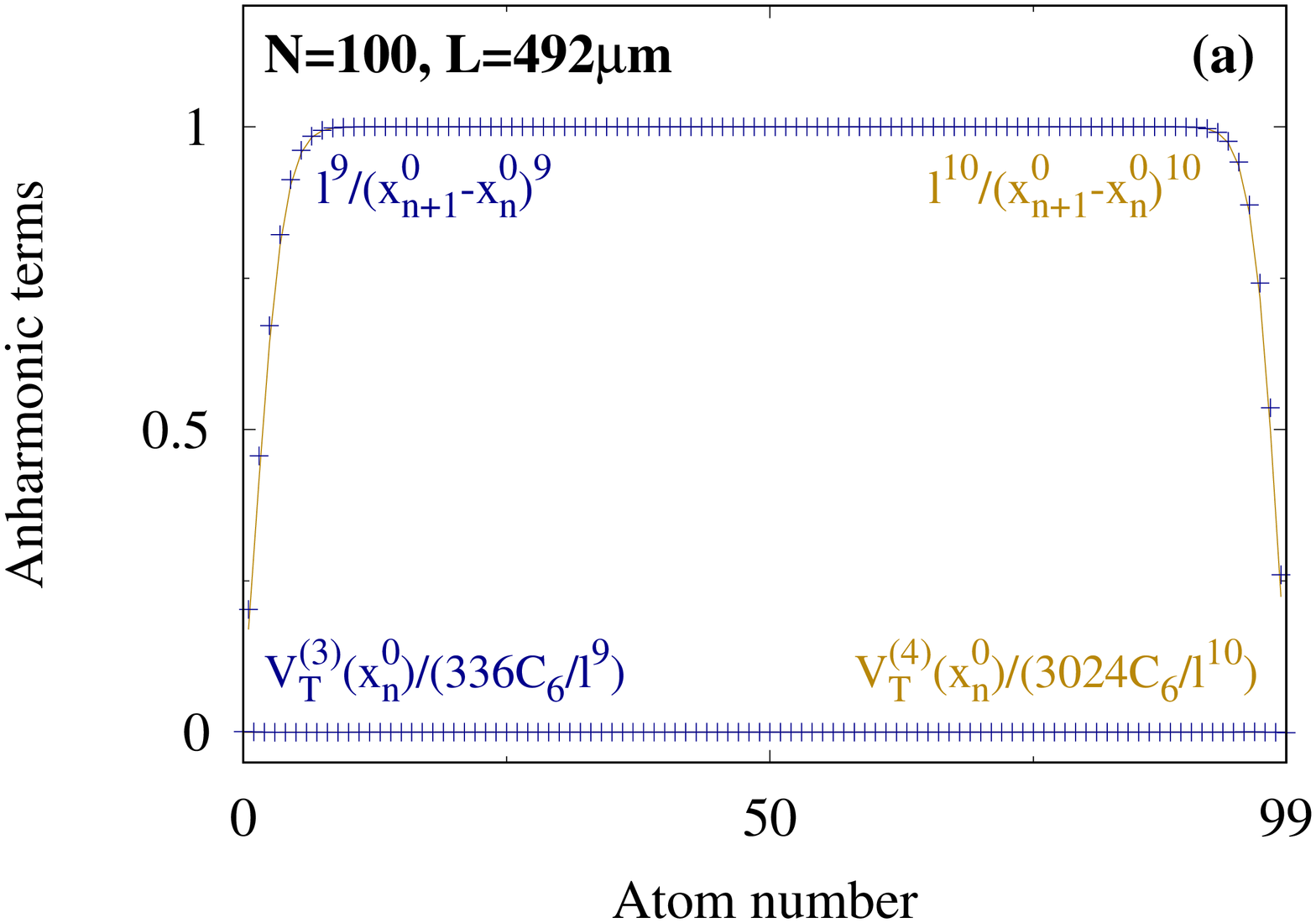}
    \caption{\label{fig:SM:anharm}
      Third-- (blue crosses) and fourth--order (golden lines)
      anharmonic amplitudes from the interaction energy
      (non--zero amplitudes)
      and the trapping potential (negligible amplitudes),
      calculated for the shorter chain of Fig.~3 in the main
    text, just before the first expulsion.
    They are expressed in units of
    their bulk values,
    namely
    $56C_6/l^9$ and $126C_6/l^{10}$.
  }
\end{figure}

The leading anharmonic contribution to the Hamiltonian
follow from the 
third-- and fourth--order terms in the displacements $\{u_n\}$.
For gases, they yield two--body collisions
which are essentially instantaneous. By contrast, for Rydberg chains,
they generate many--body correlations over the characteristic time
$\tau_\mathrm{propag}$
for propagation along the chain, set by the
sound velocity.
They are mostly due to interactions 
and occur in the chain bulk, where their probability does not
depend on position (see Fig.~\ref{fig:SM:anharm}).
They are much less probable near the edges,
where the trapping potential
leads to larger distances between the static equilibrium positions
of the atoms.

The role of these anharmonic processes  is twofold.
First, they are responsible for
thermalization and ergodicity
on a timescale involving $\tau_\mathrm{propag}$.
Second, they set
the (longer) timescale ensuring the adiabaticity of the
compression between two atomic expulsions.
The classical--dynamics simulations reported
in Ref.~\cite{nguyen:PRX2018} have
shown that, for the shorter chain of Fig.~3 in the main text
($N_i=100$),
compression rates of the order of $40\,\mathrm{\mu m/ms}$
are adequate. The optimal compression rate
will be investigated elsewhere.

For gases, anharmonic processes directly
drive the atomic expulsions,
which immediately
follow two--atom collisions during which one atom
has acquired enough energy.
Their relation to expulsions is more involved for Rydberg chains.
If the trap size is such that an expulsion is expected
($T\rightarrow\infty$), ergodicity 
causes the system to explore various
configurations until the leftmost atom is expelled with the energy $V_L$.
If no expulsion is expected
($T$ finite), the compression of the trap causes an increase in energy
due to the atoms on the edges of the chain being set in motion towards
the bulk.
Expelling the leftmost atom before thermalization has
taken place (i.e.\ with an energy $>V_L$)
is likely to involve
a two--atom collision at the open end of the trap. There,
anharmonic terms are strongly suppressed (see Fig.~\ref{fig:SM:anharm}),
so that these higher--energy expulsions are rare. Instead,
the energy increase is most often mediated, through harmonic vibrations,
to the chain bulk where thermalization occurs.
The rare cases in which the leftmost atom is expelled are not captured
by our thermodynamic model. However, they are not a hindrance as long as
their rate remains small: instead, they speed up the evaporation process
with respect to our thermodynamic prediction.
The presence of a single open
end (the left end on Fig.~1 of the main text)
is favorable for two reasons: \textit{(i)} it leads to longer
propagation times and, hence, more efficient thermalization;
\textit{(ii)} it helps reduce the rate of non--thermalized expulsions.

\section{The partition function}
\begin{figure*}
  \begin{minipage}{.35\linewidth}
    \includegraphics[angle=-90,width=\linewidth]
    {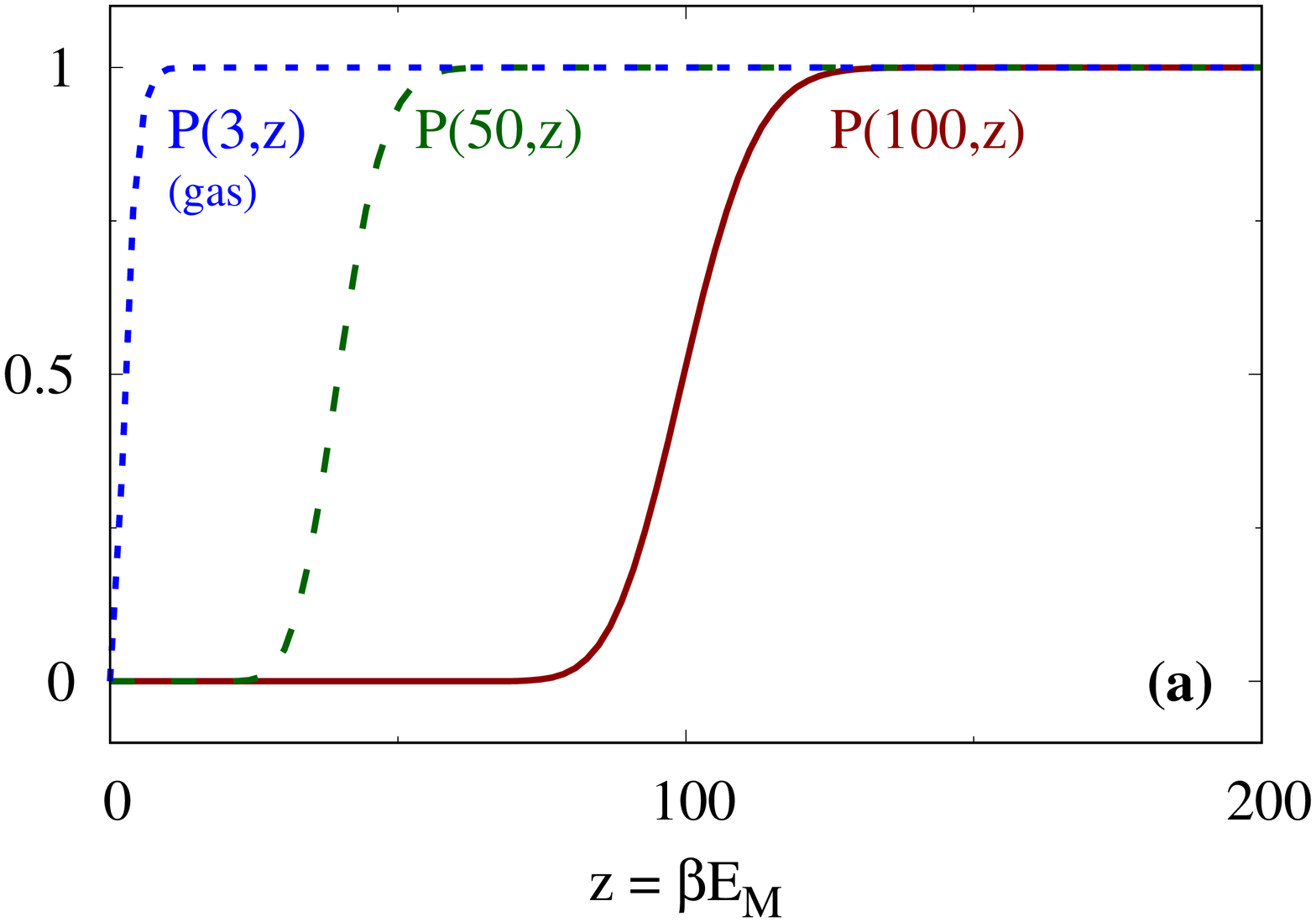}
  \end{minipage}
  \begin{minipage}{.3\linewidth}
    \vspace*{-.2cm}
    \includegraphics[angle=-90,width=\linewidth]
    {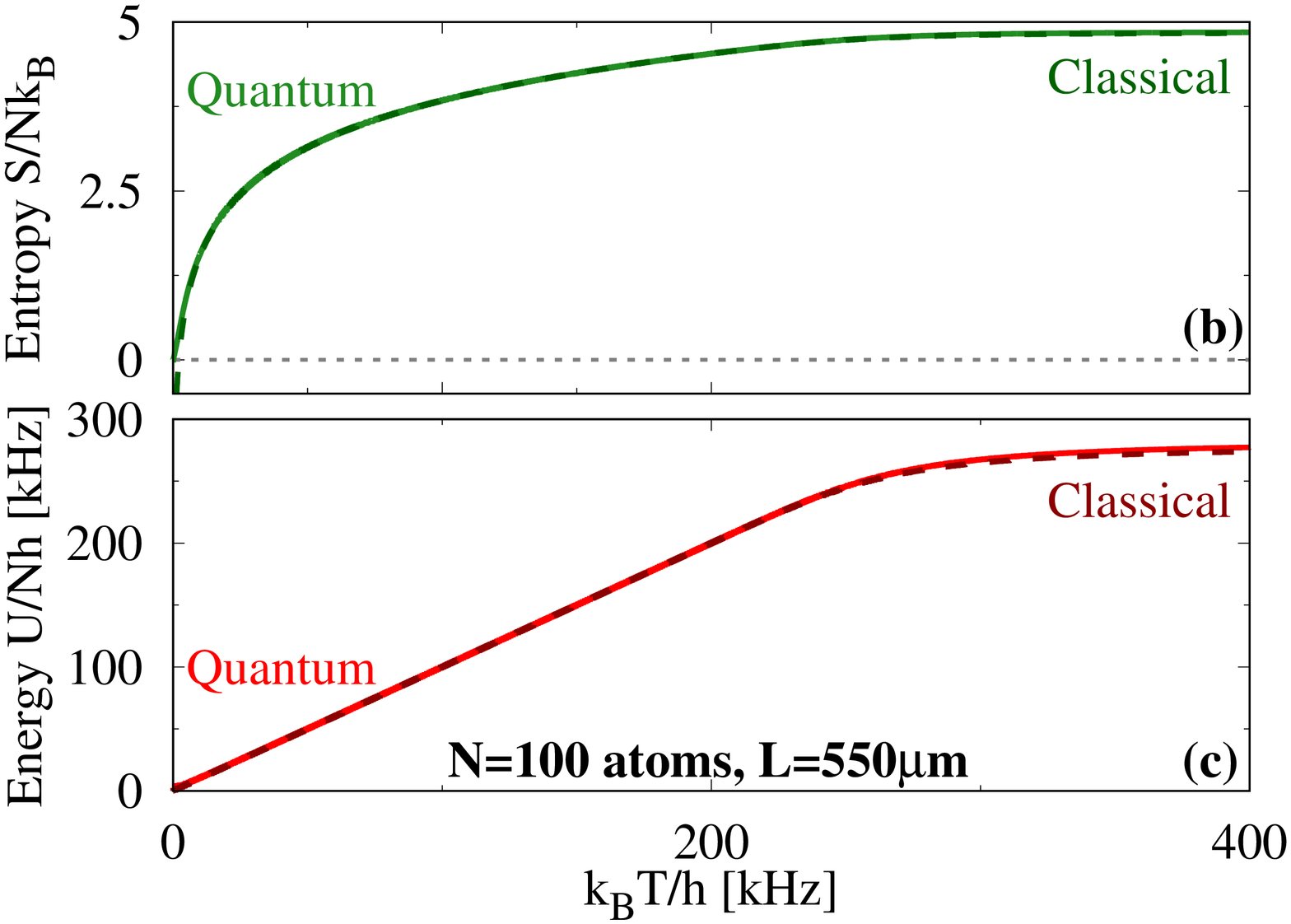}
  \end{minipage}
  \begin{minipage}{.3\linewidth}
    \vspace*{-.2cm}
    \includegraphics[angle=-90,width=\linewidth]
    {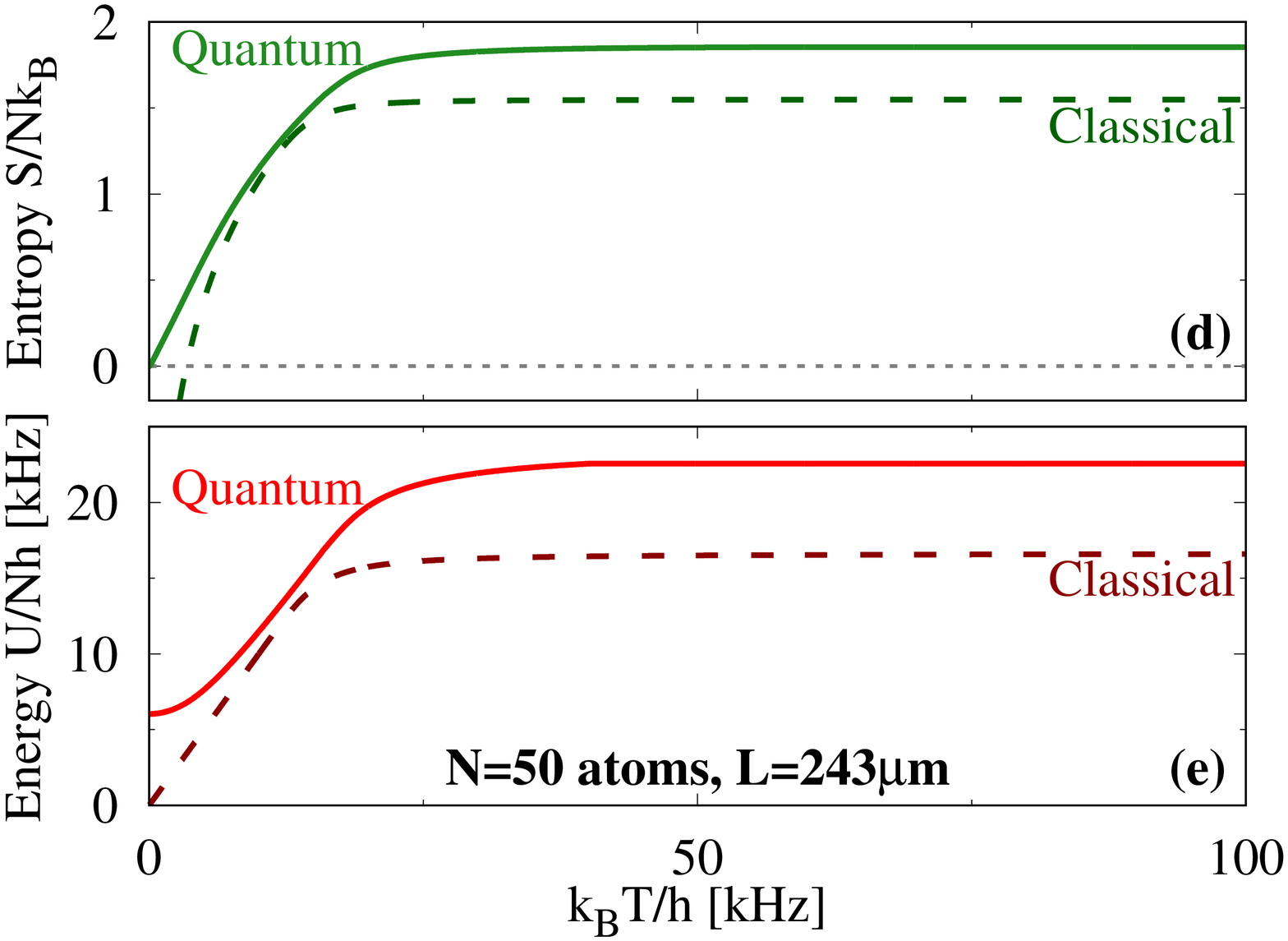}
  \end{minipage}
  \caption{\label{fig:SM:thermo}
    \textit{(a)} Normalized incomplete Gamma function $P(a,z)$
    as a function of $z=\beta E_M$, for $a=3$, $50$, and $100$.
    \textit{(b)} and \textit{(c)}: $s=S/N$ and
    $u=U/N$  as a function of $T$
    for $100$ atoms in a trap of size $L=550\,\mathrm{\mu m}$
    (beginning of the evaporation on Fig.~3 of the
    main text). \textit{(d)} and \textit{(e)}: $s$ and $u$
    for $50$ atoms with
    $L=243\,\mathrm{\mu m}$ (close to the end of the evaporation
    on Fig.~2).
    }
\end{figure*}

\emph{Normalized lower incomplete Gamma function} ---
The thermodynamics of the (classical or quantum)
truncated Boltzmann distribution
involve the normalized lower incomplete Gamma function
$P(a,z)$, defined as \cite{NIST:DLMF}:
\begin{equation}
  \label{eq:SM:incomplete_Gammafun}
  P(a,z) \: = \:
  \frac{\gamma(a,z)}{\Gamma(a)}
  \: = \:
  \frac{1}{\Gamma(a)}
  \int_0^a dt\: e^{-t} t^{z-1}
  \ .
\end{equation}
For given values of the trap size $L$ and atom number $N$,
the classical partition function
$Z^\mathrm{cl}$ is proportional to $P(N,\beta E_M)/\beta^N$. Hence,
$a$ is of the order of $N$, whereas
$z=\beta E_M$ is the ratio of the
threshold energy to the temperature.
For a given $a$, the function $P(a,z)$ resembles a step function
(see Fig.~\ref{fig:SM:thermo}\textit{(a)}) which
is equal to
$0$ for small $z$ (representing the truncation for large $T$)
and to
$1$ for large $z$ (truncation plays no role for small $T$).
The smooth transition occurs for $z\approx a$, so that
truncation plays a  role for 
$\kB T/E_M \gtrsim 1 /a$. The parameter $a=3$ 
for a gas 
in a truncated 3D harmonic trap \cite{luiten:PRA1996}, whereas
for Rydberg chains $a\approx N$ ranges from $40$ to $1000$.
Hence, Rydberg chains are affected by the truncation
starting from much lower temperatures than gases are.

\emph{Quantum partition function} ---
For a given $L$, and 
assuming  $E_M^\mathrm{quant}\gg E_{\mathrm{ZP}}+\hbar\omega_N$,
we evaluate the quantum partition function
$Z^\mathrm{quant}$ for $\kB T\gtrsim E_M^\mathrm{quant}/N$
using Eq.~2 in the main text,
replacing $E_M^\mathrm{cl}$ by
$E_M^\mathrm{quant}$. 
We go beyond the quasiclassical integral expression and
include the leading--order quantum correction, 
proportional to $\hbar^2$ \cite[\S 33]{landau5:Elsevier1980}.
Hence, we write 
$Z^\mathrm{quant}=Z^\mathrm{cl}(1+\braket{\hbar^2\chi_2})$,
where 
the correction $\braket{\hbar^2\chi_2}$
is expressed in terms of the moments
$\braket{x_k^2}_\mathrm{cl}$, $\braket{p_k^2}_\mathrm{cl}$
and $\braket{x_k^2p_k^2}_\mathrm{cl}$ of 
$Z_\mathrm{cl}$. We find:
\begin{multline}
  \braket{\hbar^2\chi_2} \:
  E_M^2/\sum_{k=1}^N(\hbar\omega_k)^2=
  \\
  \frac{z^2}{24}
  \left(
    -1+    [3z-5(N+1)]
    \frac{z^Ne^{-z}}{\Gamma(N+2)}
    \frac{1}{P(N,z)}
  \right)
  \ ,
\end{multline}
with $z=\beta E_M$.
For $\kB T< E_M/N$, we use
the quantum partition
function $Z_0=\prod_{k=1}^N[\csch(\beta\hbar\omega_k/2)/2]$
of a non--truncated 
chain.
The functions
$U^\mathrm{quant}$ and $S^\mathrm{quant}$
overlap with those extracted from $Z_0$
for a range of values of $\kB T$,  
thus yielding the full quantum thermodynamic functions.
The classical and quantum predictions
for $U$ and $S$ are compared on
Fig.~\ref{fig:SM:thermo}.
At the beginning of the evaporation (panel \textit{(b)}),
they only differ over a narrow range of temperatures near $T=0$;
the difference is more striking near the end of the evaporation
(panel \textit{(c)}).

\emph{Numerical evaluation} ---
The evaluation of
$U(L,T)$ and $S(L,T)$ involves calculating $P(a,z)$
for 
$40\leq a\leq 1000$. 
In order to capture the steep variation of these functions for
$z\sim a$, we resort to arbitrary--precision numerics
using
the Boost.Multiprecision C++ library \cite{boost:multiprecision}.

\section{ \label{sec:SM:constantN}
  Constant atom number plateaux}
\begin{figure}
  \centering
  \includegraphics[angle=-90,width=.8\linewidth]
  {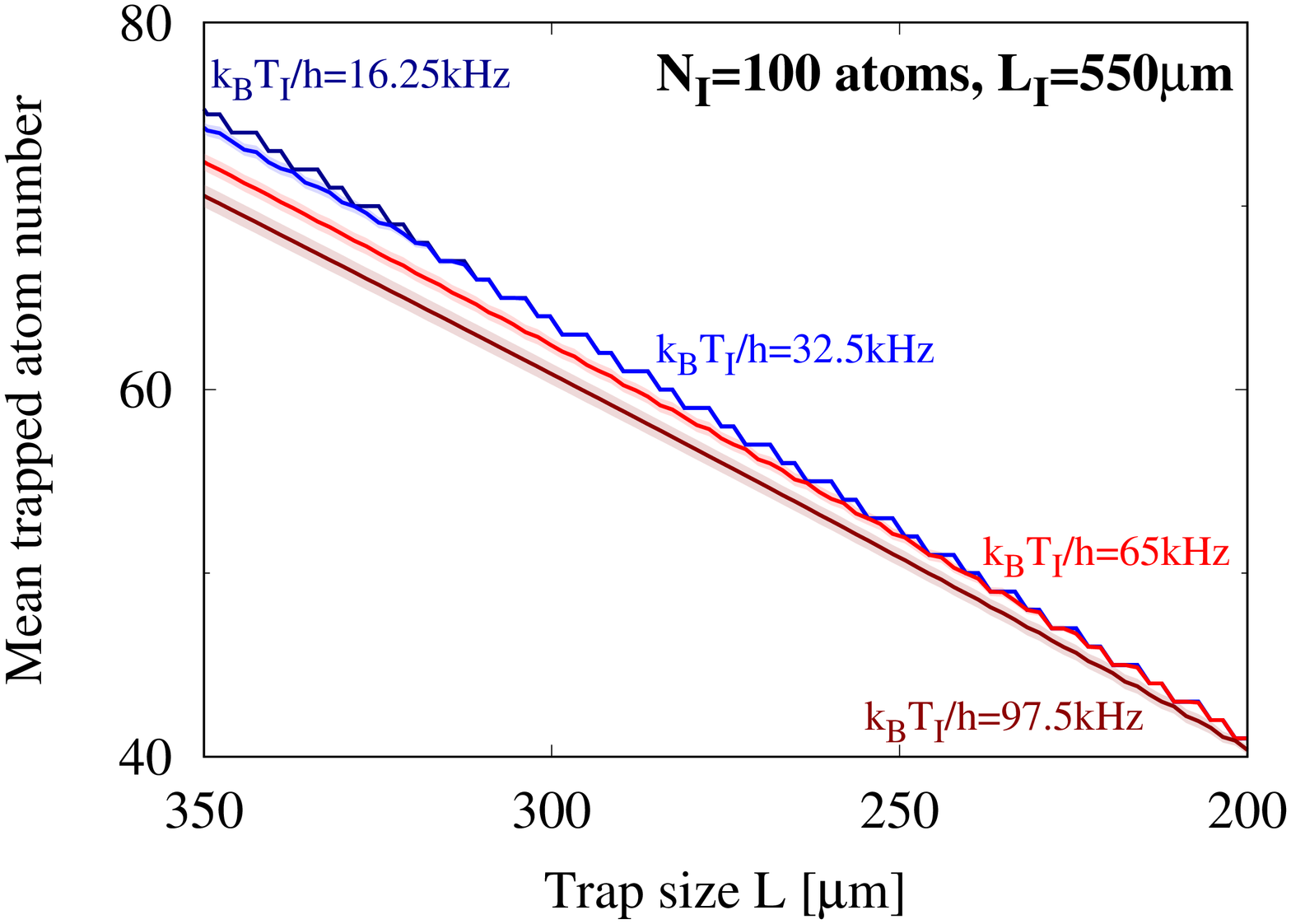}
  \caption{\label{fig:SM:constantNplateaux}
    Mean trapped atom number as a function of the trap size $L$
    for the short chain initially comprising $N_I=100$ atoms with
    $L_I=550\,\mathrm{\mu m}$ for various initial temperatures. The red
    curve, calculated for $\kB T_I=65\,\mathrm{kHz}$, corresponds to Fig.~4
    of the main text. The discrete steps on the atom number become visible
    at the end of the evaporation; for lower initial temperatures, they are
    well resolved earlier on.
    The shaded areas show the mean standard deviation due to the
    initial fluctuations $\Delta U_I=U_I/\sqrt{N_I}$ on the
    quadratic energy.}
\end{figure}
Between two atomic expulsions, the chain undergoes an adiabatic
compression during which $N$ remains constant
(see Fig.~3 in the main text).
For 
$\kB T_I/h\sim 65\,\mathrm{kHz}$,
these constant--$N$ plateaux are smoothed out for most of the evaporation
because of
the uncertainty $\Delta U_I=U_I/\sqrt{N_I}$ on the initial
energy. Indeed, it reflects on the entropy as
$\Delta S_I=\Delta U_I/T_I$, and leads to
sizable fluctuations $\Delta S^{(N)}$
during most of the evaporation. 
These yield 
the
uncertainty $\Delta L_f^{(N)}=\Delta S^{(N)}/S_\mathrm{max}^{(N)\prime}(L_f)$
on the trap size $L_f^{(N)}$ at which the atom $N$ is expelled.

For shorter chains,
the constant--$N$
plateaux become well resolved
at the end of the evaporation,
as the chain approaches its ground state. 
For Fig.~4 in the main text and
$\kB T_I/h=65\,\mathrm{kHz}$, these plateaux
are visible when the remaining trapped atom number  $N\lesssim 45$
(see Fig.~\ref{fig:SM:constantNplateaux}),
in agreement with the classical--dynamics results of
Ref.~\cite{nguyen:PRX2018}. The plateaux are resolved earlier on for lower
initial temperatures and later on for higher ones.

\section{Quasi--universality}
We now focus on longer chains with $N\sim 1000$ 
and $l\sim 5\,\mathrm{\mu m}$. Then,
the quadratic energy
$U(N,L,T)=Nu(l,T)$, the entropy $S(N,L,T)=Ns(l,T)$,
their maxima
$U_\mathrm{max}(N,L)=Nu_\mathrm{max}(l)$ and
$S_\mathrm{max}(N,L)=Ns_\mathrm{max}(l)$, and the zero--point energy
$E_\mathrm{ZP}(N,l)=Ne_\mathrm{ZP}(l)$, are all extensive.

\emph{Energy and entropy  ---}
We consider two consecutive adiabatic plateaux corresponding to $N$
and $N-1$ trapped atoms.
Equation~5 in the main text provides the initial energy per particle
$u_i^{(N-1)}=U_i^{(N-1)}/(N-1)$ for the second plateau in terms of
its final value for the first one, $u_f^{(N)}=U_f^{(N)}/N$, and the
mean distance $l_f^{(N)}=L_f^{(N)}/N$:
\begin{equation}
  \label{eq:SM:uiNm1}
  u_i^{(N-1)}=u_f^{(N)}+(u_f^{(N)}+7C_6/l_f^{(N)6}-V_L)/N
  \ .
\end{equation}
Hence, starting from the first atomic expulsion, 
$u$
remains close to the universal curve $u=u_\mathrm{max}(l)$,
within small deviations which decrease like $1/N$.
Furthermore, the entropies $s^{(N)}=s_\mathrm{max}(l_f^{(N)})$, which are
constant during each plateau, all lie near the universal curve
$s=s_\mathrm{max}(l)$. Both of these properties are illustrated on
Fig.~5 of the main text for various 
$u_I=U_I/N_I$, which set the mean atomic distance $l_1$ at which
the first expulsion occurs.

\begin{figure*}
  \begin{minipage}{.45\linewidth}
    \centering
    \includegraphics[angle=-90,width=.9\linewidth]
    {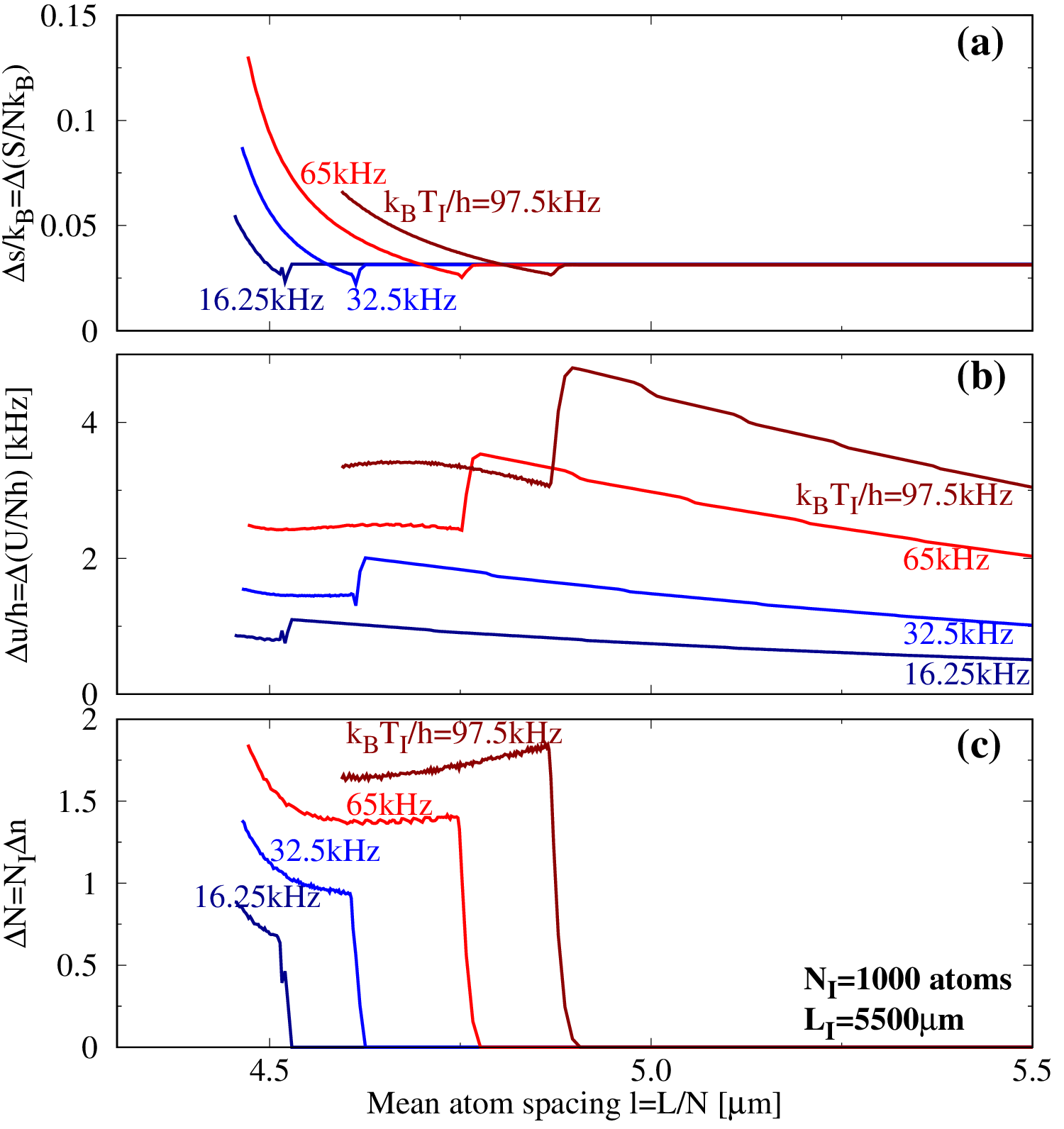}
  \end{minipage}
  \begin{minipage}{.45\linewidth}
    \centering
    \includegraphics[angle=-90,width=.9\linewidth]
    {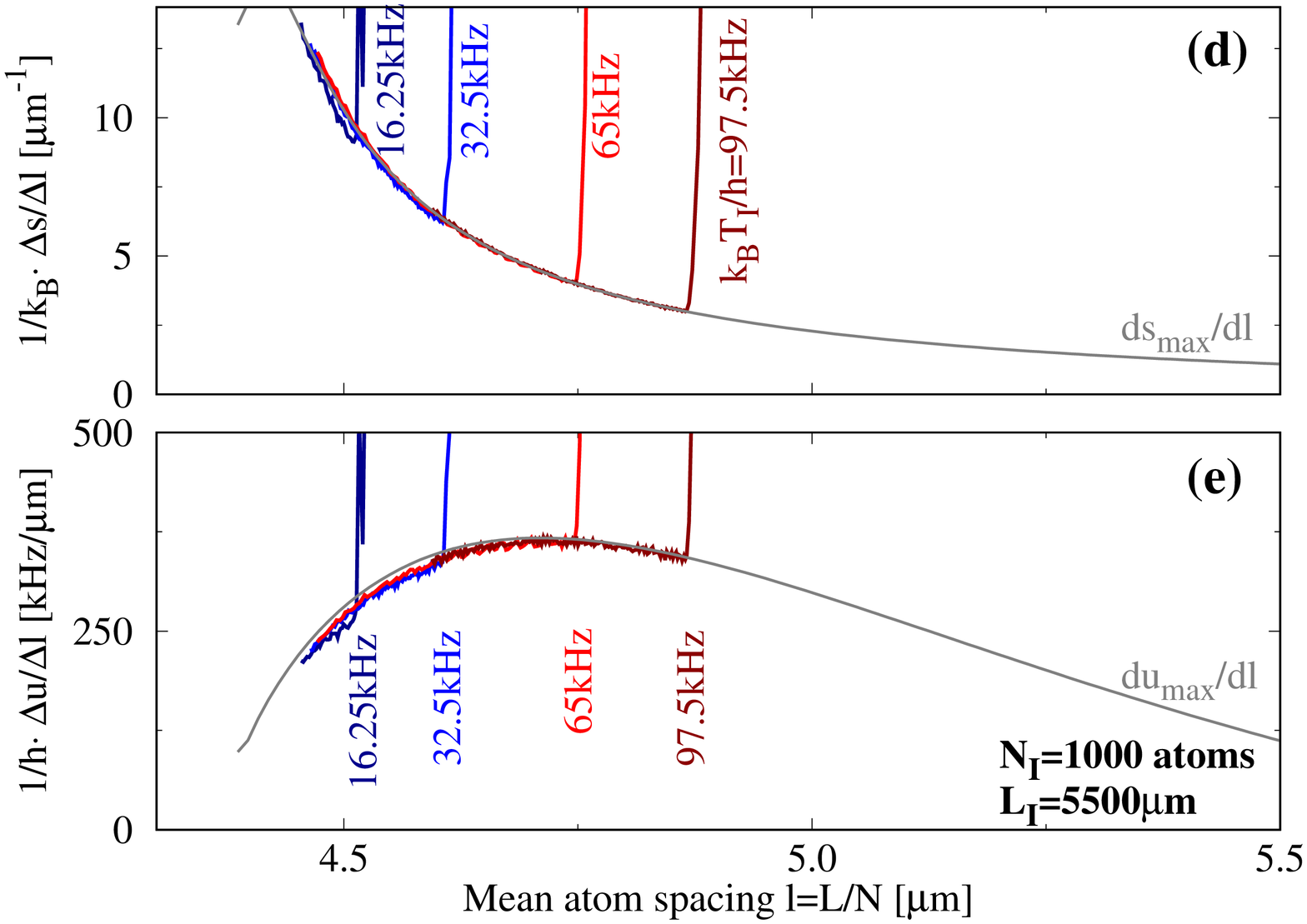}
  \end{minipage}
  \caption{
    \label{fig:SM:standarddev}
    Standard deviations \textit{(a)} $\Delta s$,
    \textit{(b)} $\Delta u$,
    and \textit{(c)} $\Delta n$ on the entropy per particle $s=S/N$,
    the energy per particle $u=U/N$, and
    the remaining atom fraction $n=N/N_I$, for the long chain
    of Fig.~5 in the main text.
    The ratios \textit{(d)} $\Delta s/\Delta l$ and
    \textit{(e)} $\Delta u/\Delta l$,
    with $\Delta l=(l/n)\Delta n$, closely follow the derivatives
    $ds_\mathrm{max}/dl$ and $du_\mathrm{max}/dl$ starting from the first
    expulsion.}
\end{figure*}
\emph{Fluctuations ---}
The quasi--universality 
of the evaporation
constrains the fluctuations $\Delta u$ and $\Delta s$
on the energy and entropy per particle to follow
those on the atomic distance, $\Delta l$.
Neglecting the small deviations
from the universal curves $u=u_\mathrm{max}(l)$ and $s=s_\mathrm{max}(l)$,
they satisfy 
$\Delta u/\Delta l=u_\mathrm{max}'(l)$ and
$\Delta s/\Delta l=s_\mathrm{max}'(l)$
(see Fig.~\ref{fig:SM:standarddev}).

The constraint on $\Delta u/\Delta l$ has an important consequence.
As $l$ decreases, 
$u_\mathrm{max}(l)$ tends towards
$e_\mathrm{ZP}(l)$
(see Fig.~5\textit{(b)} in the main text). Hence, the
derivative $u_\mathrm{max}'(l)$ goes to zero. The fluctuations $\Delta u$
do not vanish, therefore $\Delta l$ increases and so does
$\Delta n=(n/l)\Delta l$ (see Fig.~\ref{fig:SM:standarddev}\textit{(a)}).
Thus, as long as the quasi--universal regime holds, the constant--$N$
plateaux will be poorly resolved. If the evaporation proceeds further,
it will eventually drive the system
out of the universal regime. Then, we expect to recover the short--chain
behavior described in Sec.~\ref{sec:SM:constantN}. For the chain considered
in Fig.~\ref{fig:SM:standarddev}, this occurs beyond the validity
range of our assumption
$E_M^\mathrm{quant}\gg E_\mathrm{ZP}+\hbar\omega_N$, and will be
investigated elsewhere.

\begin{figure*}
  \begin{minipage}{.325\linewidth}
    \includegraphics[angle=-90,width=.9\linewidth]
    {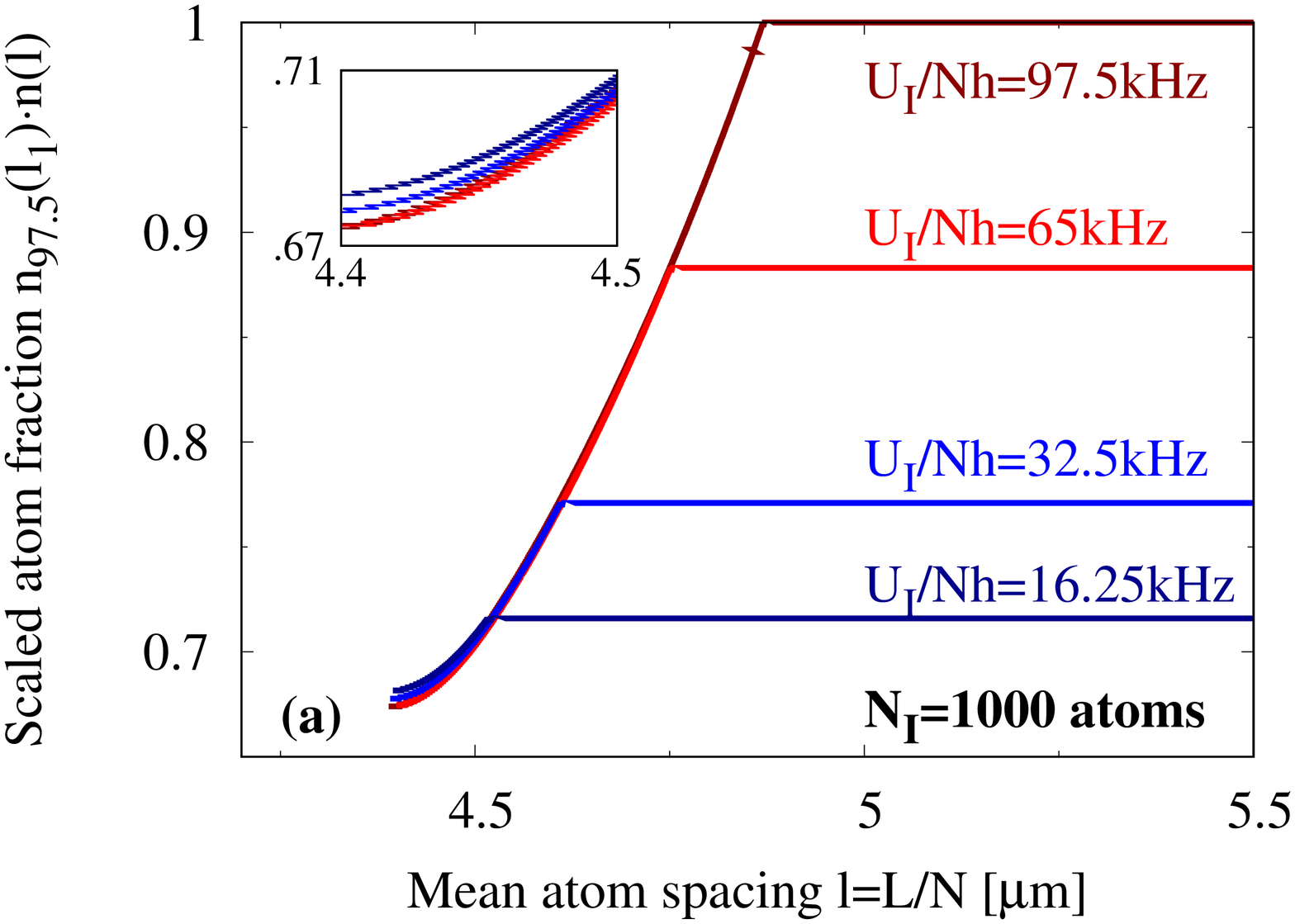}
  \end{minipage}
  \begin{minipage}{.325\linewidth}
    \includegraphics[angle=-90,width=.9\linewidth]
    {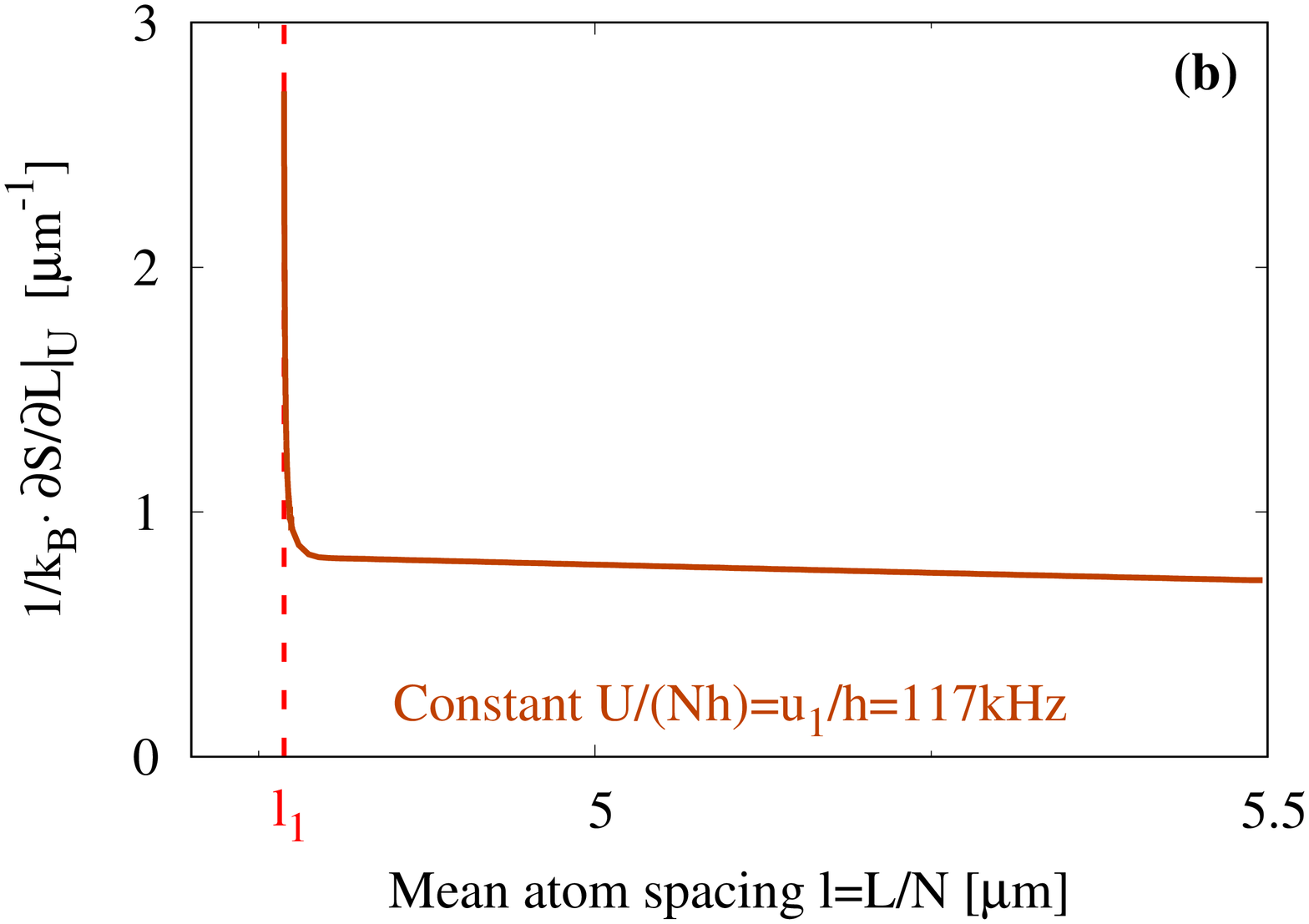}
  \end{minipage}
  \begin{minipage}{.325\linewidth}
    \includegraphics[angle=-90,width=.9\linewidth]
    {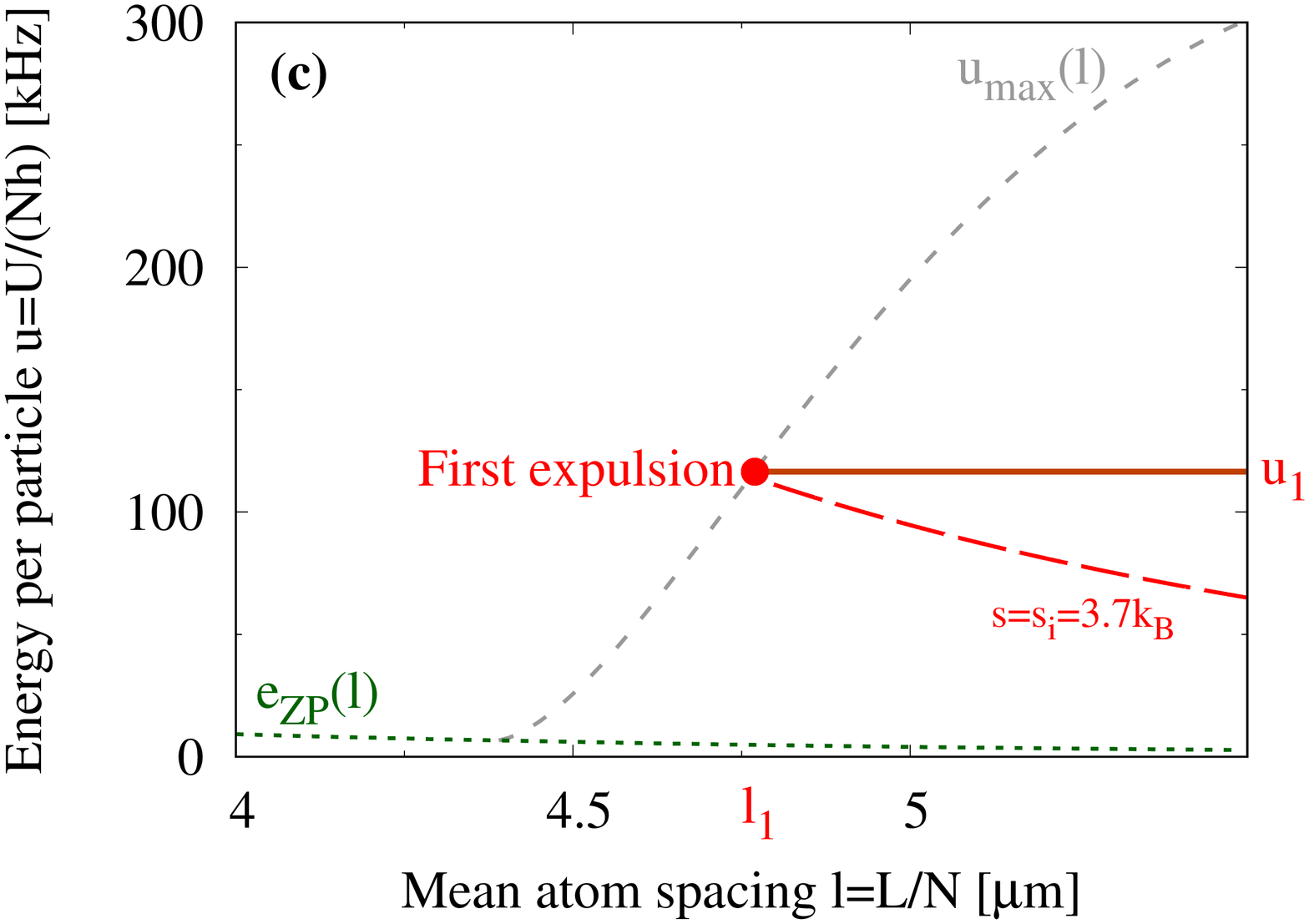}
  \end{minipage}
  \caption{
    \label{fig:SM:atomfrac_nonUniv}
    \textit{(a)} Scaled atom fraction $n=N/N_I$ 
    for $N_I=1000$ and various $u_I$, showing
    an approximate scaling whose breakdown is visible in the inset.
    \textit{(b)} The derivative $\partial s/\partial l|_u$ calculated
    along the horizontal dark red path on panel \textit{(c)}.
    This path
    crosses the curve $u=u_\mathrm{max}(l)$ (dashed gray line) at the
    first expulsion point  $(l_1,u_1)$ for $u_I/h=65\,\mathrm{kHz}$.
    The dotted--dashed red line 
    shows the isentropic curve followed up to the first expulsion.
  }
\end{figure*}
\emph{Non--universality of $N/N_I$ ---}
The entropy per particle $s(l,u)$ may be seen as a function of
$l$ and $u$.
The derivative
$\partial s/\partial u|_l=1/T$
goes to zero on the curve $u=u_\mathrm{max}(l)$,
which is reached for $T\rightarrow\infty$. However, our numerical
results show that $\partial s/\partial l|_u$ diverges
along the curve $u=u_\mathrm{max}(l)$ (see Fig.~\ref{fig:SM:atomfrac_nonUniv}).
Therefore, $s(l,u)$ may not be linearized near this curve,
and the entropy difference
$s^{(N-1)}-s^{(N)}=s(l_i^{(N-1)},u_i^{(N-1)})-s(l_f^{(N)},u_f^{(N)})$
goes to zero slower than $1/N$. This rules out
any exact universal behavior for the atom number fraction $n=N/N_I$.
However, the deviation from universality is small.
For a given $N_I$, we consider two initial energies $u_{I1}<u_{I2}$,
and compare the curves $n_{u_{I1}}(l)$ and $n_{u_{I2}}(l)$ for $l<l_1$,
where $l_1$ is the mean atom spacing leading to the first expulsion
for $u_{I1}$. Our numerical results show that
these two curves nearly satisfy the scaling relation which would
have been exact had $\partial s/\partial l|_u$ not been divergent,
namely $n_{u_{I2}}(l_1)n_{u_{I1}}(l)\approx n_{u_{I2}}(l)$
(see Fig.~\ref{fig:SM:atomfrac_nonUniv}\textit{(a)}, whose inset
highlights the breakdown of this scaling behavior).

The divergence of $\partial s/\partial l|_u=p/T$ along the curve
$u=u_\mathrm{max}(l)$
signals that the pressure $p$ goes to infinity faster than $T$ does.
This starkly contrasts with the behavior of the
ideal gas, where $p/T=n\kB$
is finite, its constant value being set by the particle density $n$.

%\bibliography{rydbergevapbib}

\end{document}